\documentclass[epjST]{svjour}
\usepackage[applemac]{inputenc}
\usepackage{graphics}
\usepackage[table]{xcolor}
\usepackage{graphicx}
\usepackage{bm}
\usepackage{bbm}
\usepackage{physics}
 \usepackage{braket}
    \usepackage{url}
    \usepackage{bbding}
\usepackage{pifont}
\usepackage{color}
\usepackage{cite}
\usepackage{amsmath}
\usepackage{amssymb}
\usepackage{wrapfig}
\usepackage{ragged2e}
\usepackage{comment}
\begin{document}
\title{Odd-frequency superconducting pairing in one-dimensional systems}
\author{Jorge Cayao,\thanks{\email{jorge.cayao@physics.uu.se}} Christopher Triola, \and Annica M. Black-Schaffer}
\institute{Department of Physics and Astronomy, Uppsala University, Box 516, S-751 20 Uppsala, Sweden}
\abstract{
\justify
Odd-frequency superconductivity represents a truly unconventional ordered state which, in contrast to conventional superconductivity, exhibits pair correlations which are odd in relative time and, hence, inherently dynamical. In this review article we provide an overview of recent advances in the study of odd-frequency superconducting correlations in one-dimensional systems. In particular, we focus on recent developments in the study of nanowires with Rashba spin-orbit coupling and metallic edges of two-dimensional topological insulators in proximity to conventional superconductors. These systems have recently elicited a great deal of interest due to their potential for realizing one-dimensional topological superconductivity whose edges can host Majorana zero modes. We also provide a detailed discussion of the intimate relationship between Majorana zero modes and odd-frequency pairing.  
Throughout this review, we highlight the ways in which odd-frequency pairing provides a deeper understanding of the unconventional superconducting correlations present in each of these intriguing systems and how the study and control of these states holds the potential for future applications.
} 

\maketitle

\section{Introduction}
\label{sec0}
Since its discovery in 1911\cite{kamerlingh1911resistance}, superconductivity has had a profound impact on our world, leading to applications such as magnetic resonance imaging, precision magnetic field measurements using SQUIDs, and precision voltage measurements using Josephson junctions. Characterized by zero electrical resistance and the complete expulsion of magnetic flux below a critical temperature, this unique phase of matter has its origin in the macroscopic phase coherence of electrons due to the formation of Cooper pairs. In the conventional BCS theory of superconductivity, the order parameter, $\Delta$, characterizing the superconducting phase can be thought of as a many-body wavefunction describing these electron pairs. Therefore, the fermionic nature of the constituent electrons in the electron pair imposes fundamental constraints on the symmetries of the order parameter, namely that it must be antisymmetric under the interchange of all quantum numbers associated with the constituent electrons, including both spin and spatial degrees of freedom.
This leads to the conventional symmetry classification of Cooper pairs as either spin-singlet and even in spatial parity or spin-triplet and odd in spatial parity. The first class includes the $s$-wave pairing in conventional BCS superconductors \cite{PhysRev.108.1175}, as well as the $d$-wave pairing in high temperature superconductors \cite{RevModPhys.72.969}. The second class includes $p$-wave and $f$-wave pairing proposed in some unconventional superconductors \cite{RevModPhys.75.657}. 
 
We stress that the above classification scheme assumes that the interaction is instantaneous and, thus, the electrons pair at equal times. However, in more realistic extensions to BCS theory, which account for retarded interactions due to the exchange of bosonic modes\cite{eliashberg1960interactions,abrikosov2012methods,mahan2013many}, the quantity $\Delta$ can acquire a time-dependence through expectation values of the form $\Delta(t)\sim\langle \psi(t)\psi(0)\rangle$, where $\psi(t)$ annihilates an electronic state at time $t$. To account for this possibility of unequal-time pairing the order parameter may be written in terms of time-ordered propagators, which we also refer to as pair amplitudes, describing the dynamics of Cooper pairs, from which we can obtain more general symmetry constraints. In this case, the pair amplitude must be antisymmetric under the exchange of all the quantum numbers of the electrons plus the exchange of the relative time coordinates. In this extended classification scheme, if the pair amplitude is even in the relative time or, equivalently the relative frequency, $\omega$, then the amplitude must fall into one of the above mentioned classes: even-$\omega$, spin-singlet, and even-parity (ESE) or even-$\omega$, spin-triplet, and odd-parity (ETO). However, the pairing function could also be \textit{odd} in frequency, in which case Cooper pairs could be odd-$\omega$, spin-triplet, and even-parity (OTE) or odd-$\omega$, spin-singlet, and odd-parity (OSO). Due to their odd time-dependence, odd-$\omega$ pair amplitudes vanish at equal times, and hence odd-$\omega$ pairing is an intrinsically dynamical phenomenon. These symmetry classes are summarized in Table \ref{tab1} with the even-$\omega$ classes in black and the odd-$\omega$ in red.

\renewcommand{\arraystretch}{1.7}
\begin{table}
\label{tab1}
\begin{center}
 \begin{tabular}{|c | c | c | c|| c |} 
 \hline 
 & Frequency & Spin & Space & {\bf Total}\\ [0.5ex] 
\hline\hline
ESE & even & singlet & even & odd \\ [1pt]
 \hline
{\color{red}OSO} & odd & singlet & odd &odd \\[1pt]
 \hline
 ETO & even & triplet & odd&odd \\[1pt]
 \hline
{\color{red}OTE} & odd & triplet & even &odd \\[1pt]
 \hline
 \end{tabular}
 \end{center}
 \caption{Classification scheme for Cooper pair amplitudes due to the constraints given by Fermi-Dirac statistics when the relevant degrees of freedom are frequency, spin, and spatial coordinates. The pair amplitudes may be either even or odd under the exchange of any one of these indices (relative time, spin, and spatial coordinates), but they must always be odd under the simultaneous exchange of all indices.}
 \end{table}
 
The possibility of odd-$\omega$ pairing was first envisaged by Berezinskii to describe superfluid $^{3}$He with OTE symmetry \cite{bere74}. It was later extended to superconductors with OTE \cite{PhysRevLett.66.1533,PhysRevB.46.8393} and OSO pairing \cite{PhysRevB.45.13125,PhysRevB.47.513}. In these initial discussions, the odd-$\omega$ pairing represented the dominant pairing channel giving rise to a bulk order parameter, $\Delta(t)$, with odd $t$-dependence. Although there is no experimental confirmation of this intrinsic effect, it motivated a spur of subsequent studies searching for odd-$\omega$ pairing not as a thermodynamically stable phase but as an effect accompanying conventional even-$\omega$ order parameters \cite{PhysRevB.46.10812,PhysRevB.48.7445,PhysRevB.47.6157,PhysRevLett.70.2960,BALATSKY1994363,PhysRevB.49.8955,PhysRevB.52.1271,PhysRevB.52.15649,PhysRevLett.74.1004,PhysRevLett.74.1653,0953-8984-9-2-002,PhysRevB.59.R713,PhysRevB.60.3485,doi:10.1143/JPSJ.69.2229,PhysRevB.64.132507,doi:10.1143/JPSJ.72.2914,PhysRevB.77.144513,PhysRevB.79.132502,PhysRevB.79.174507,doi:10.1143/JPSJ.78.123710,PhysRevB.83.140509,doi:10.1143/JPSJ.80.054702,doi:10.1143/JPSJ.80.044711,doi:10.1143/JPSJ.81.033702,doi:10.1143/JPSJ.81.123701,PhysRevB.85.174528,PhysRevB.85.224509,doi:10.1143/JPSJS.81SB.SB015,PhysRevLett.110.107005,doi:10.7566/JPSJ.82.104702,PhysRevLett.112.167204,PhysRevB.90.115154,doi:10.7566/JPSJ.83.123704,PhysRevLett.115.036404,KASHIWAGI201629,PhysRevB.93.224511}. The basic idea behind many of these proposals is the use of a conventional BCS superconductor with a static order parameter, $\Delta$, as a source for Cooper pairs whose odd-$\omega$ dynamics are subsequently induced by altering some of the properties of the electronic states.

One of the most promising routes for realizing odd-$\omega$ pairing in superconducting systems is offered by heterostructures, in which the electronic properties of a conventional even-$\omega$ superconductor can be modified by the presence of different, usually non-superconducting, materials. 
Perhaps the simplest example of such a heterostructure is the normal-superconductor (NS) junction, in which it has been shown that the breaking of translation symmetry at the interface generically leads to the transmutation of even-parity amplitudes to odd-parity, while preserving the spin structure of the Cooper pairs, thus giving rise to odd-$\omega$ pairing \cite{PhysRevB.71.094513,PhysRevB.72.140503,PhysRevLett.98.037003,PhysRevLett.99.037005,Eschrig2007,PhysRevB.76.054522}. Alternatively, if there is a spin-active field, e.g.~from a ferromagnet (F), then the even-$\omega$ spin-singlet pair amplitudes of a conventional superconductor can be converted to a odd-$\omega$ spin-triplet amplitude, leading to the creation of spin-triplet correlations with robust $s$-wave spatial symmetry in ferromagnet-superconductor (FS) junctions. 
Following this mechanism, equal-spin OTE pairing has been proposed in diffusive FS junctions with inhomogeneous magnetization and conventional spin-singlet $s$-wave superconductors \cite{PhysRevLett.86.4096,Kadigrobov01}. 
This equal-spin OTE pairing offers an explanation for the observed long-range penetration of Cooper pairs into diffusive F regions \cite{longrangeExp,PhysRevB.58.R11872,Keizer06,PhysRevLett.104.137002,wang10,Robinson59,PhysRevB.82.100501,Cirillo_2017,PROSHIN2018359,PhysRevB.97.100502}. Additionally, a growing body of evidence indicates that such spin-triplet pairs emerge in FS junctions \cite{BELZIG19991251,PhysRevB.62.11377,PhysRevB.62.11846,PhysRevLett.86.308,PhysRevB.68.064513,PhysRevLett.88.047003,PhysRevLett.90.137003,Fominov2003,RevModPhys.77.935,RevModPhys.77.1321,PhysRevLett.98.077003,PhysRevLett.98.107002,PhysRevB.75.134510,PhysRevB.98.161408}, including the paramagnetic Meissner effect \cite{PhysRevB.64.132507,PhysRevB.64.134506,PhysRevX.5.041021} and zero-bias peaks \cite{PhysRevB.92.014508,bernardo15}. 
These advances in understanding how odd-$\omega$ equal-spin triplet pairing emerges in FS junctions are fascinating not just for their fundamental importance in unconventional superconductivity \cite{RevModPhys.63.239,Maeno94,PhysRevLett.80.3129,PhysRevLett.107.077003} but also for their potential applications in superconducting spintronics \cite{LinderNat15,7870d3ff91ed485fa3e55e901ff81c80,EschrigNat15,0034-4885-78-10-104501}.

Odd-$\omega$ pairing has also been shown to possess a deep relationship with topological superconductors, an exciting class of systems attracting much attention for their potential to host Majorana zero modes (MZMs) \cite{PhysRevLett.100.096407,PhysRevB.79.161408,PhysRevLett.105.077001,PhysRevLett.105.177002}. MZMs are their own antiparticles with non-Abelian statistics, making them hold great potential for applications in topological quantum computation \cite{kitaev,RevModPhys.80.1083,Sarma:16}. 
Two of the most promising classes of systems proposed to exhibit MZMs are heterostructures made by combining conventional $s$-wave superconductors with either: nanowires with Rashba spin-orbit coupling (SOC) or the metallic edges of two-dimensional topological insulators (2D TIs), resulting in one-dimensional (1D) systems with MZMs at the system boundaries. Due to their strong intrinsic SOC, large degree of tunability, and 1D nature, an enormous amount of experimental activity has focused on these and similar systems during the past years, where also initial evidence of MZMs has been reported \cite{Aguadoreview17,LutchynReview08,zhangreview,tkachov19review}. In order to reach the topological superconducting phase, various configurations have been proposed for these heterostructures, usually in the presence of a magnetic field. In general, the combination of all these ingredients, SOC, magnetism, and conventional superconductivity, gives rise to a variety unconventional superconducting correlations with exotic spin textures. 
Motivated by the promise of 1D topological superconductivity and its MZMs, much effort has been dedicated to investigating the proximity effect in these and related systems. These efforts have also shed light on the origin of odd-$\omega$ spin-triplet correlations in topological insulators \cite{PhysRevB.86.075410,PhysRevB.86.144506,PhysRevB.87.220506,PhysRevB.92.205424,Lu_2015,PhysRevB.92.100507,PhysRevB.96.155426,PhysRevB.96.174509,bo2016,PhysRevB.97.075408,PhysRevLett.120.037701,PhysRevB.97.134523} and nanowires with Rashba spin-orbit coupling \cite{PhysRevB.92.134512,Ebisu16,PhysRevB.95.184518,PhysRevB.98.075425,PhysRevB.99.184501} in proximity to conventional spin-singlet $s$-wave superconductors. It is now understood that SOC can induce odd-$\omega$ spin-triplet correlations also in the absence of magnetism. Moreover, it is becoming clear that odd-$\omega$ spin-triplet pairing can be the dominant channel in topological superconductors, that there is a close relationship between MZMs and odd-$\omega$ pairing, and that the odd-$\omega$ dependence of superconducting correlations can have profound effects on these systems. 

\begin{figure}
  \begin{center}
\includegraphics[width=.75\textwidth]{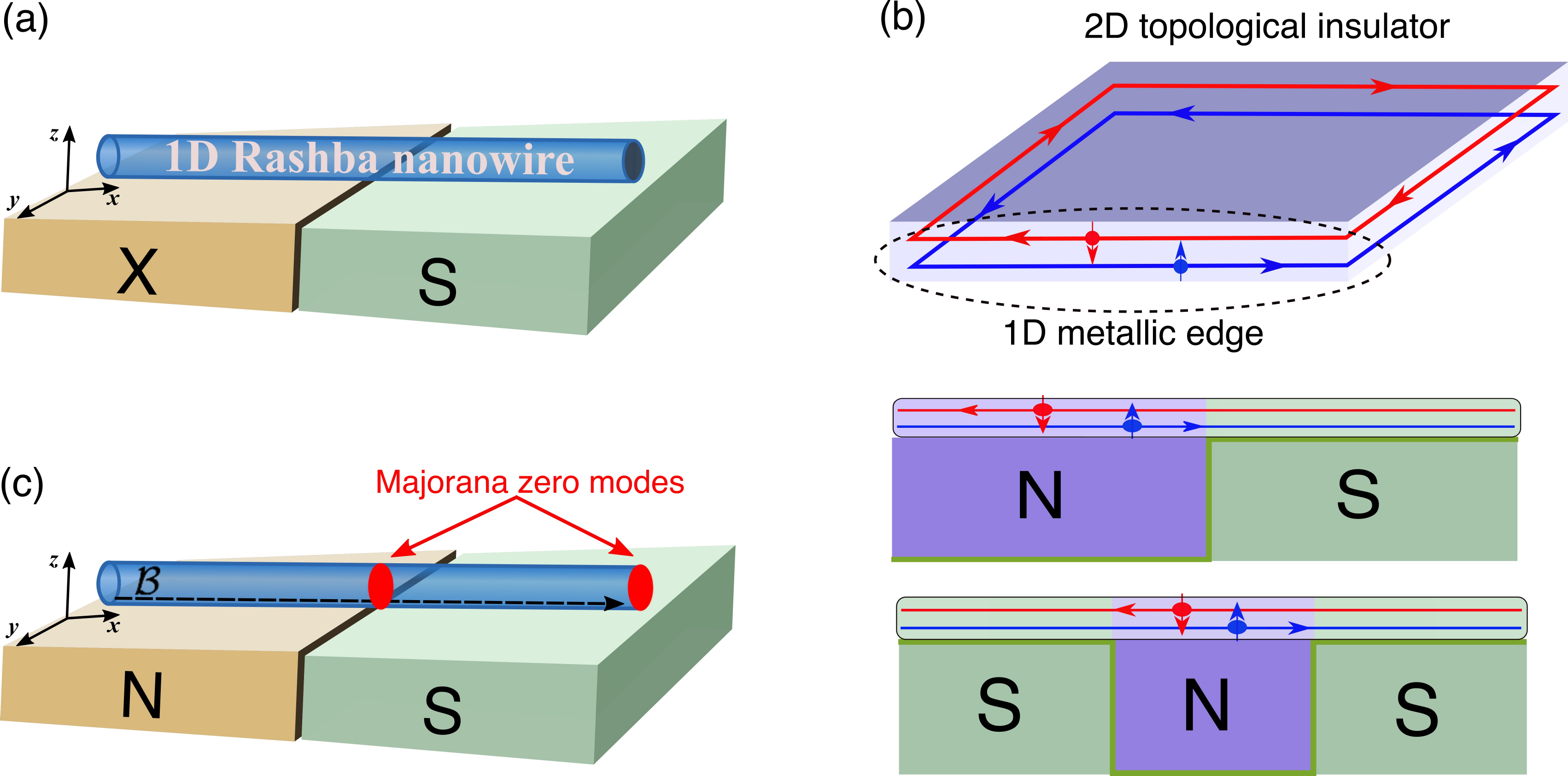} 
  \end{center}
  \caption{Sketch of 1D systems with SOC which exhibit OTE pairing and are promising for 1D topological superconductivity. 
  (a) The right region of a nanowire with Rashba SOC is placed in contact with a conventional spin-singlet $s$-wave superconductor (S), while its left region  is in contact with either a normal (X=N) or superconducting lead (X=S). (b) A 2D TI is characterized by having an insulating bulk and a 1D metallic edge with counter-propagating  modes carrying opposite spin (top, blue and red filled circles with arrows indicating spin direction). The 1D metallic edge is placed in proximity to a superconductor S to form NS (middle) and SNS (bottom) junctions. The profile of $\Delta$ is indicated by green lines. (c) Under the right conditions, the systems in (a,b) exhibit a topological superconducting phase with MZMs (red filled circles) at the boundary between regions of different topology.
Reprinted modified figure with permission from [J. Cayao and A. M. Black-Schaffer, Phys. Rev. B 96, 155426 (2017); Phys. Rev. B 98, 075425 (2018)] Copyright (2017) and (2018) by The American Physical Society.
  }
  \label{Sketchsystem}
  \end{figure}

In this brief review we discuss recent advances in the field of induced odd-$\omega$ pairing in 1D systems. In particular, we focus on those systems most relevant for 1D topological superconductivity and the realization of MZMs: nanowires with Rashba SOC and the metallic edges of 2D TIs, shown schematically in  Fig.\,\ref{Sketchsystem}. In Section \ref{sec1}, we discuss how the constraints imposed by Fermi-Dirac statistics lead directly to a classification of superconducting symmetries which explicitly allows for odd-$\omega$ pairing. In Section \ref{sec2}, we illustrate how odd-$\omega$ pairing can arise in 1D NS and SNS junctions without any spin-active fields. Then, in Sections \ref{sec3} and \ref{sec4}, the role of Rashba SOC and the helical nature of 2D TIs are described, respectively. Specifically, we discuss how the spin-active aspects of these two systems allow the generation of novel superconducting correlations, including odd-$\omega$ spin-triplet pairing. In Section \ref{sec5}, we examine the relationship between odd-$\omega$ pairing and MZMs, illustrating that MZMs are always accompanied by odd-$\omega$ pair amplitudes. Finally, in Section \ref{sec7}, we present a few concluding remarks.

Due to the necessary focus of this review, there are a number of systems we inevitably have to omit in our discussion. Apart from the FS and NS heterostructures mentioned above, induced odd-$\omega$ correlations have also been shown to appear in double quantum dots coupled to conventional superconductors \cite{PhysRevB.90.220501,PhysRevB.93.201402}, enabled by the dot index which acts as an additional degree of freedom, and in superconductors subjected to a time-dependent drive \cite{PhysRevB.94.094518,triola17}. Notably, odd-$\omega$ pairing has also been predicted to emerge in bulk multiband superconductors, where the additional band degree of freedom plays a crucial role in the symmetry classification \cite{PhysRevB.88.104514,PhysRevB.92.094517,PhysRevB.92.224508}.  For more information on odd-$\omega$ pairing in FS junctions we refer to Refs.\,\cite{RevModPhys.77.1321,Samokhvalov_2016}, in NS junctions to Refs.\,\cite{Golubov2011x,Tanaka2018}, and for its relation with topology to Ref.\,\cite{Nagaosa12}. For a current overview of the field, we also refer to Ref.\,\cite{Balatsky2017}.

\section{Odd-$\omega$ pairing and Fermi-Dirac statistics}
\label{sec1}
The key feature of a superconductor is the presence of a charged superfluid composed of pairs of electrons, Cooper pairs, which form below a critical temperature due to electron-electron interactions.  
As with other ordered states, the properties of a superconducting state depend intimately on the symmetry of the superconducting order parameter, $\Delta$. As mentioned in the introduction, within BCS theory and its generalizations, the superconducting order parameter is directly related to the anomalous propagator, $f$, $\Delta\sim f$, where $f$ describes the dynamics of the Cooper pairs. Therefore, the symmetries of $f$ determine the nature of the superconducting state. The anomalous propagator can be defined as  $f(t)= -i\langle T\psi(t) \psi(0)\rangle$, where $\psi(t)$ annihilates a electron at time $t$, and $T$ is the time-ordering operator, making its interpretation as a pair amplitude obvious. It is also similar to the normal propagator, or Green's function, describing the propagation of free electrons, $g(t)\equiv -i\langle T\psi(t)\psi^\dagger(0) \rangle$. The behavior of both the normal (electron-electron) and anomalous (electron-hole) propagators or Green's functions \cite{zagoskin} can be obtained from their Matsubara, time-ordered, or the retarded and advanced representations. Throughout this article we will work with the retarded and advanced propagators, unless otherwise specified. 

In systems with only a single relevant band/orbital, the normal and anomalous Green's functions depend only on the time coordinates (or frequency), spins, and spatial degrees of freedom of the electrons. However, it is often convenient to group these two propagators into a single Nambu space Green's function \cite{zagoskin}
\begin{equation}
\label{GF}
G^{r}(x,x',\omega)=
\begin{pmatrix}
G^{r}_{ee}&G^{r}_{eh}\\
G^{r}_{he}&G^{r}_{hh}
\end{pmatrix}\,,
\end{equation}
where $G^{r}_{ee}$ and $G^{r}_{eh}$ correspond to the normal (electron-electron) and anomalous (electron-hole) components, which are, in general, $2\times 2$ matrices in spin space making $G^{r}$ a $4\times4$ matrix. In this general case, we have $f^{r}_{\sigma\sigma'}(x,x';\omega)=[G_{eh}^{r}(x,x',\omega)]_{\sigma\sigma'}$. We note that, typically, pair amplitudes, $f^{r}$, are complex numbers, therefore, throughout this work we make a distinction between the pair amplitude and the pair magnitude: $|f^{r}|=\sqrt{f^{r}(f^{r})^{*}}$.

Considering the definition of the time-ordered pair amplitudes, it can easily be shown that Fermi-Dirac statistics imposes the following constraint
\begin{equation}
f^{t}_{\sigma\sigma'}(x,x';\omega)=-f^{t}_{\sigma'\sigma}(x',x;-\omega)\,,
\label{eq:constraint_t}
\end{equation}
where $f^{t}$ is the time-ordered anomalous propagator written in the frequency domain. This antisymmetry condition leads directly to the symmetry classification in Table \ref{tab1}, which is a complete description of the symmetries of superconducting correlations when the only degrees of freedom are time, spins, and position. To make contact with the retarded and advanced correlators used in this review, we note that the above constraint is equivalent to the following relation 
\begin{equation}
f^{r}_{\sigma\sigma'}(x,x';\omega)=-f^{a}_{\sigma'\sigma}(x',x;-\omega)\,,
\label{eq:constraint}
\end{equation}
where we have used the advanced counterpart in passing to negative frequencies \cite{Balatsky2017}. 

It is often useful to decompose the pair amplitude into different terms depending on their spin symmetry. In the basis given by $(\psi_{\uparrow},\psi_{\downarrow},\psi_{\uparrow}^{\dagger},\psi_{\downarrow}^{\dagger})$ we find that a generic pair amplitude has the form
\begin{equation}
\label{decomposeSPIN}
f^{r}(x,x',\omega)=(f^{r}_{0}\sigma_{0}+f^{r}_{j}\sigma_{j})i\sigma_{y}\,,
\end{equation}
where $\sigma_{i}$ is the $i^{\text{th}}$ Pauli matrix in spin space and repeated indices imply summation. Note that, in this basis, $f^{r}_{0}$ and $f^{r}_{3}$ correspond to the spin-singlet and mixed spin-triplet configurations ($\uparrow\downarrow\mp\downarrow\uparrow$), respectively, both possessing spin-projection $S_{z}=0$. In contrast, $if^{r}_{2}\mp f^{r}_{1}$ correspond to the equal spin-triplet configurations ($\uparrow\uparrow/\downarrow\downarrow$) with spin-projection $S_{z}=\pm1$. 
Given this decomposition in spin space, we see that the antisymmetry condition in Eq. (\ref{eq:constraint}) implies
\begin{equation}
\label{pairantisymmetry}
\begin{split}
f^{r}_{0}(x,x';\omega)&=f^{a}_{0}(x',x;-\omega)\,,\\
f^{r}_{i}(x,x';\omega)&=-f^{a}_{i}(x',x;-\omega)\,,\quad i=1,2,3\, .
\end{split}
\end{equation}
Notice the absence of an overall minus sign on the right-hand side of the first expression due to the antisymmetry of the $i\sigma_y$ matrix under the exchange of spin indices. The symmetry classes of each of these amplitudes is then uniquely specified by analyzing either the frequency or spatial dependencies of the components.

For instance, if the spin-singlet (S) amplitude $f^{r}_{0}(x,x';\omega)$ is even for $\omega\rightarrow-\omega$ (and $f^r\rightarrow f^a$) then it must be even under the exchange of the spatial coordinates $x\leftrightarrow x'$, and we say that the pairing has even-$\omega$ spin-singlet even-parity symmetry, corresponding to the ESE class in Table \ref{tab1}. 
On the other hand, if the spin-triplet (T) amplitude $f^{r}_{i}(x,x';\omega)$ is even for $\omega\rightarrow-\omega$ (and $f^r\rightarrow f^a$) then it must be odd under the exchange of $x\leftrightarrow x'$, hence the pairing has even-$\omega$ spin-triplet odd-parity symmetry and corresponds to the ETO class in Table \ref{tab1}.
However, it is possible that the pair amplitude $f^{r}_{0,i}(x,x';\omega)$ acquires a minus sign for $\omega\rightarrow-\omega$ (and $f^r\rightarrow f^a$). In this case, $f^{r}_{0}(x,x';\omega)$ must be odd under the exchange of spatial coordinates, giving rise to odd-$\omega$ spin-singlet odd-parity pairing in the OSO symmetry class. 
Similarly, $f^{r}_{i}(x,x';\omega)$ could acquire a minus sign for $\omega\rightarrow-\omega$ (and $f^r\rightarrow f^a$) implying that it is even under the exchange spatial coordinates, leading to pair correlations with odd-$\omega$ spin-triplet even-parity symmetry in the OTE class. Thus we see that Fermi-Dirac statistics permits exactly four symmetry classes (ESE, OSO, ETO, OTE) when the pair amplitude of Cooper pairs depends on frequency and the spins and positions of the paired electrons. In the remainder of this review we will examine various concrete examples of systems in which these four symmetry classes can be realized. 

\section{Odd-$\omega$ pairing in 1D NS and SNS junctions}
\label{sec2}
While the systems relevant for topological superconductivity usually possess some non-trivial spin-dependence, odd-$\omega$ pairing can already emerge at interfaces between superconductors and normal metals in the absence of any spin-dependent effects. Before going further it is useful to show this effect in spin-degenerate 1D NS and SNS junctions, which represent two of the simplest geometries with interesting relations to transport observables. 

We note that odd-$\omega$ pairing in 1D was initially investigated within the quasiclassical Usadel and Eilenberg frameworks \cite{PhysRevB.71.094513,PhysRevB.72.140503,PhysRevLett.98.037003,PhysRevLett.99.037005,Eschrig2007,PhysRevB.76.054522}. However, in this section we discuss results based on a fully quantum mechanical approach, where retarded and advanced Green's functions are calculated from scattering processes that satisfy outgoing wave boundary conditions \cite{PhysRev.175.559}. 
In the quasiclassical approach the Fermi energy is usually assumed to be the largest energy scale in the system and fast oscillating terms with the Fermi wave vector drop off as  these oscillations take place on length scales much shorter than the those relevant to the problem \cite{larkin69}. However, the Fermi energy is not always the largest energy scale and, importantly, we are often interested in observing interference effects which are purely quantum mechanical and, hence, hard to address within the quasiclassical approach.
The method employed here is simple and experimentally relevant as it includes processes, such as Andreev reflection (AR), that can be observed e.g.~in conductance measurements. Moreover, AR processes, a property specific to NS interfaces \cite{Andree64}, are well understood to describe the core physics underlying the proximity effect \cite{PhysRevB.25.4515,Pannetier2000,Klapwijk2004}. In the following subsections we highlight the relationship between AR and odd-$\omega$ pairing, which provides useful intuition for understanding the 1D systems discussed later in this review.

\subsection{NS junctions}
In this subsection, we consider a ballistic semi-infinite NS junction whose interface is located at $x=0$. The superconducting region S, $x>0$, corresponds to a spin-singlet $s$-wave superconductor, like Al or Nb, and is characterized by having a finite order parameter $\Delta$, while the normal region N, $x<0$, has $\Delta=0$. The Bogoliubov-de Gennes (BdG) Hamiltonian describing this NS junction, in the basis $(\psi_{\uparrow},\psi_{\downarrow}^{\dagger})$, is given by
\begin{equation}
\label{BdG2}
H_{\rm 1D}(x)=\Big(\frac{p_{x}^{2}}{2m}-\mu(x)+V(x)\Big)\tau_{z}+{\rm Re}(\Delta(x))\tau_{x}+{\rm Im}(\Delta(x))\tau_{y}
\end{equation}
where $p_{x}=-i\hbar\partial_{x}$, $\Delta(x)=\theta(x)\Delta\,{\rm e}^{i\phi}$, $\mu(x)=\mu_{\rm N}\theta(-x)+\mu_{\rm S}\theta(x)$ is the chemical potential profile, $\phi$ the superconducting phase, $V(x)$ models scattering at the interface, and $\tau_{i}$ represents the $i^{\text{th}}$ Pauli matrix in electron-hole subspace. In NS junctions $\phi$ does not play any role and can be gauged away; however, as we will see, it plays an essential role in SNS systems.

\begin{figure}
  \begin{center}
\includegraphics[width=.8\textwidth]{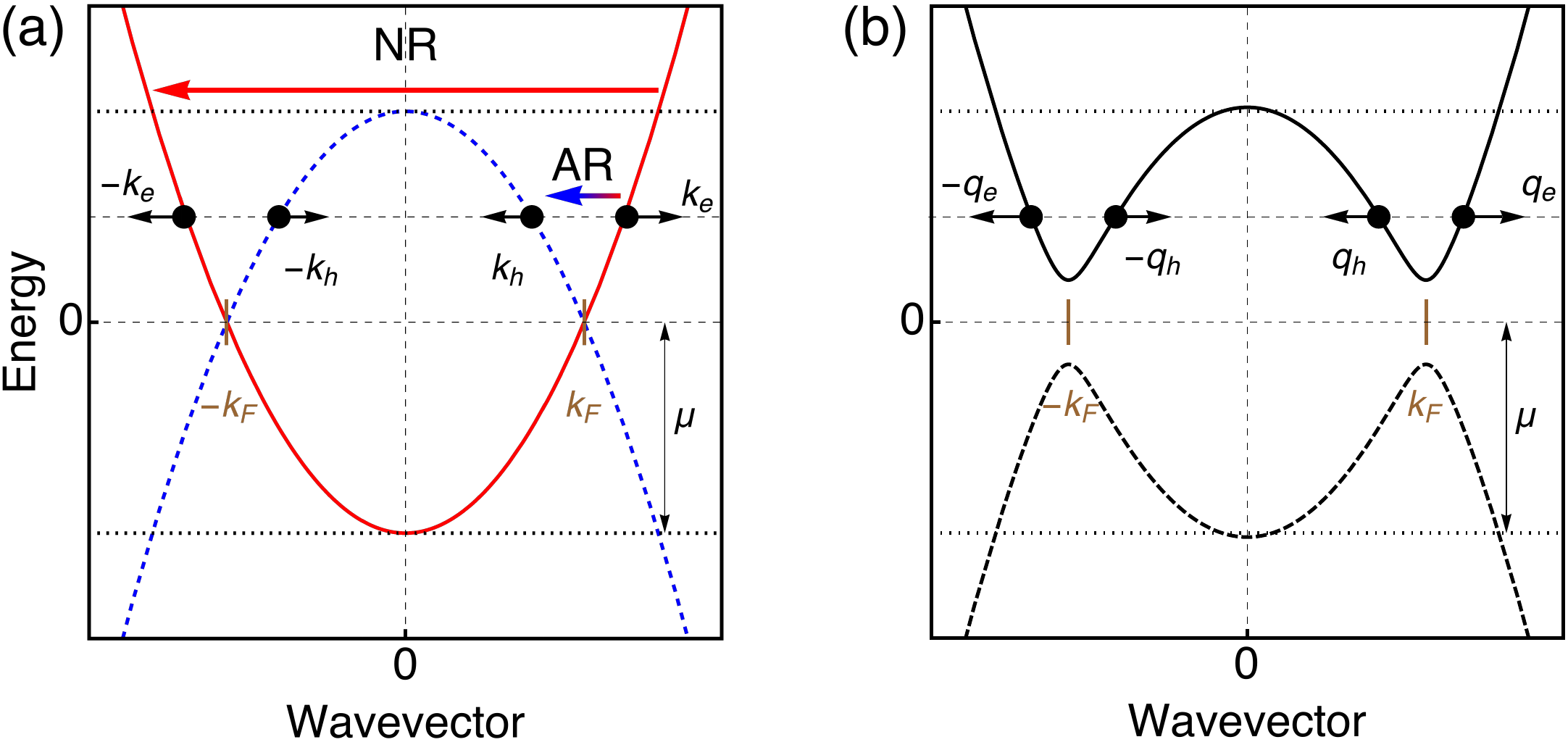} 
  \end{center}
\caption{(a) Energy-momentum dispersion in a normal metal (N) with $\Delta=0$, where solid red (dashed blue) parabola corresponds to electrons (holes). The bottom (top) of the electron (hole) band is indicated by dotted horizontal lines.
(b) A finite order parameter $\Delta\neq0$ in the superconductor (S) opens an energy gap at the Fermi momenta $\pm k_{\rm F}$  and mixes electron with hole bands, leading to quasielectron and quasihole bands with positive (solid) and negative (dashed) energies, respectively. At the NS interface four scattering processes from incident particles towards the interface occur, two from N and two from S, which allow for  the calculation of the Green's function. This can be visualized by fixing the energy to a positive value, indicated by horizontal dashed line above zero energy, where small horizontal black arrows denote the direction of motion of the states. Then,  an incident  right-moving electron from N with wave vector $k_{e}$  can be reflected back at the  NS interface  into a left-moving electron  in N with $-k_{e}$, a process known as normal reflection (NR) and depicted by solid red arrow, or into a left-moving hole in N with $k_{h}$, known as Andreev reflection (AR) and shown by red-blue arrow. Moreover, it can also be transmitted into  S  as a right moving quasielectron with wave vector $q_{e}$ or right moving quasihole with $-q_{h}$. Other processes include right moving hole from N with $-k_{h}$, left moving quasielectron (quasihole) from S with $-q_{e}$ ($q_{h}$).}
\label{NS}
   \end{figure}

In order to calculate the Green's function for the NS junction, the scattering processes at the NS interface are constructed, as discussed in Ref.\,\cite{PhysRevB.98.075425}. 
Four kinds of scattering processes can be seen by inspection of the energy versus momentum dispersion of  Eq.\,(\ref{BdG2}) and depicted in Fig.\,\ref{NS}. One kind of process involves a right-moving electron from N with wavevector $k_{e}$ (hole with $-k_{h}$) being reflected within the N region as a left-moving electron with $-k_{e}$ (hole with $k_{h}$), a process known as normal reflection (NR).
Another process involves a right-moving electron from N with $k_{e}$ (hole with $-k_{h}$) being reflected into N as a hole with $k_{h}$ (electron with $-k_{e}$), the so-called Andreev reflection (AR). Moreover, the right moving incident particles can be also transmitted into S as right-moving quasielectrons with $q_{e}$ or right-moving quasiholes with $-q_{h}$. Similar processes are also possible for left-moving quasielectrons with $-q_{e}$ (quasihole with $q_{h}$) propagating towards the interface from inside S.  
The four scattering processes outlined above, together with their conjugated counterparts, allow calculation of the full retarded Green's function in N and S, for more details see Ref.~\cite{PhysRevB.98.075425}.

Since both N and S are trivial in spin space, the anomalous Green's function, and hence the pair amplitude, only possesses a spin-singlet component. In the N region, this component is given by \cite{PhysRevB.98.075425}
\begin{equation}
\label{eq0}
f_{0}^{r}(x,x'\omega)=\frac{\eta}{2i}r_{eh}(\omega)\,{\rm e}^{-i(k_{e}x-k_{h}x')}\,,
\end{equation} 
where  $\eta=2m/\hbar^2$, $r_{eh}(\omega)$ is the AR amplitude for a right moving electron from N, $k_{e,h}=k_{\mu_{\rm N}}\sqrt{1\pm \omega/\mu_{N}}$ with $k_{\mu_{\rm N}}=\sqrt{2m\mu_{\rm N}/\hbar^{2}}$. 
Our first observation from Eq.\,(\ref{eq0}) is that, despite being non-superconducting, N acquires superconducting correlations, $f_0^r$, through the expression in Eq.\,(\ref{eq0}), a phenomenon known as the proximity effect. 
We also note the importance of the AR processes described above, as $f_{0}^{r}$ is directly proportional to the AR amplitude $r_{eh}$. Moreover, the exponential term in Eq.\,(\ref{eq0}) mixes electron and hole wavevectors $k_{e,h}$ at different positions $x,x'$, and, therefore, mixes the spatial parity. To resolve the terms with definite spatial parity we decompose the exponential term in the limit of large $\mu_{\rm N}$, when $k_{e,h}\approx k_{\mu}\big[1\pm \omega/(2\mu_{\rm N})\big]$. In this case, we obtain the two pair amplitudes
$f_{0}^{r,{\rm O}}(x,x',\omega)=-(r_{eh}\eta/2)\,{\rm e}^{-ik^{\rm N}(x+x')}{\rm sin}[k_{\mu_{N}}(x-x')]\,,
f_{0}^{r,{\rm E}}(x,x',\omega)=(r_{eh}\eta/2i)\,{\rm e}^{-ik^{\rm N}(x+x')}{\rm cos}[k_{\mu_{N}}(x-x')]\,,$
with $k^{N}=\omega k_{\mu_{\rm N}}/(2\mu_{\rm N})$, 
where $f_{0}^{r,{\rm O (E)}}$ is manifestly odd (even) under the exchange of spatial coordinates. Since both of these have spin-singlet symmetry, Fermi-Dirac statistics dictates that $f_{0}^{r,{\rm O}}$ and $f_{0}^{r,{\rm E}}$ must correspond to OSO and ESE symmetry classes, respectively.  
With these expressions and their advanced counterparts, it is also straightforward to verify the constraint from Fermi-Dirac statistics given by Eq.~(\ref{eq:constraint}). As an illustrative example, we note that the advanced counterpart of the retarded OSO amplitude is given by $f_{0}^{a,{\rm O}}(x,x',\omega)=-(r^{*}_{eh}\eta/2)\,{\rm e}^{ik^{\rm N}(x+x')}{\rm sin}[k_{\mu_{N}}(x-x')]$. Using the identity $r_{eh}^{*}(-\omega)=r_{eh}(\omega)$, we readily see that $f_{0}^{a,{\rm O}}(x,x',-\omega)=-f_{0}^{r,{\rm O}}(x,x',\omega)$, in full agreement with Eq.~(\ref{eq:constraint}). 
The analogous relation can also be verified for the ESE pairing. Note that the method discussed here assumes zero temperature, $T=0$, in which we find that the induced pairing does not decay into the N region. We can account for finite T by using the Matsubara representation where $\omega\rightarrow i\omega$, in which case the decay length into N goes as $\sim 1/T$, see e.g.~Refs.~\cite{PhysRevB.73.014503,PhysRevB.81.014517}. 

A similar analysis as above can also be carried out in the S region. Instead of writing the full expressions we discuss only the main findings, for additional details see Ref.\,\cite{PhysRevB.98.075425}. In that work, the authors found that the pair amplitudes in the S region have contributions from the interface and the bulk. As expected, the bulk term exhibits the same ESE symmetry as the superconductor in the absence of the N region. 
At the interface, however, the AR allows the coexistence of ESE and OSO pair amplitudes which exponentially decay into the bulk. On the other hand, the NR processes emerge solely with even-spatial parity and, therefore, only contribute to ESE pairing. For transparent interfaces and equal $\mu_{i}$, NRs become negligible, leaving only contributions due to AR. 

In the case of NS junctions, the coexistence of ESE and  OSO amplitudes arises due to the breaking of translation invariance at the NS interface with the order parameter being zero in N and finite in S, namely, $\Delta(x)=\theta(x)\Delta$. This relationship between spatial translation symmetry breaking and the emergence of odd-$\omega$ pairing has been investigated in other similar geometries \cite{PhysRevB.71.094513,PhysRevB.72.140503,PhysRevLett.98.037003,PhysRevLett.99.037005,Eschrig2007,PhysRevB.76.054522}, as well as the surface of a superconducting substrate in the presence of two nanowires \cite{Triola19b}, and the surfaces of 3D TIs in the presence of a spatially non-uniform order parameter \cite{PhysRevB.86.144506}. However, we stress that breaking spatial translation invariance alone does not necessarily give rise to OSO pairing \cite{Triola19}, in the cases mentioned above it emerges due to the fact that the order parameter, $\Delta(x)$, breaks the translation-invariance.
 
If the NS junction is of finite length such that $-L<x<0$ defines the N region and $x>0$ defines the S region, sharp subgap features, known as McMillan-Rowell peaks \cite{PhysRevLett.16.453}, emerge in the LDOS due to electron-hole interference effects in the N region of the junction. In Ref.\,\cite{PhysRevB.76.054522} it was shown that the ratio of the odd- and even-frequency correlations diverges at the energies of these subgap peaks, implying a strong relation between McMillan-Rowell peaks and odd-$\omega$ pairing.

Before ending this section we point out that if we consider a superconductor with spin-triplet $p$-wave symmetry (ETO) instead of a spin-singlet $s$-wave superconductor (ESE) in Eq.\,(\ref{BdG2}),
 the induced odd-$\omega$ pairing will have OTE symmetry. This was initially considered in junctions with diffusive N regions \cite{PhysRevB.71.094513,PhysRevB.72.140503,PhysRevLett.98.037003,PhysRevLett.99.037005}. Those studies were motivated by the fact that only OTE pairing is believed to penetrate into a diffusive N region, due to impurity scattering.
It is here interesting to note that the role of ARs in the generation of even- and odd-$\omega$ amplitudes in junctions with $p$-wave superconductors is the same as in junctions with $s$-wave superconductors discussed above, despite the difference in symmetries between the superconductors involved \cite{thanos2019}. On the other hand, for $p$-wave superconductors NR terms with even-spatial parity emerge and, therefore, also contribute to the OTE pairing \cite{thanos2019}. This implies that, depending on the symmetry of the superconducting region of the NS junction (ESE or ETO), NRs may or may not contribute to odd-$\omega$ pairing. 

We thus have seen that in even simple NS junctions odd-$\omega$ pairing is induced by breaking the spatial parity of Cooper pairs, which in this case occurs due to the change of $\Delta(x)$. Here, conventional $s$-wave spin-singlet  pairing is converted into $p$-wave spin-singlet pairing.  The breaking of a single symmetry of Cooper pairs thus induces two different symmetry classes,~ESE and OSO at interfaces of NS junctions. 
Moreover, it is the AR processes that mediate the coexistence of ESE and OSO pairings at the interfaces in these junctions. 

\subsection{SNS junctions}
In this subsection we discuss the pair amplitudes that emerge in superconductor-normal-superconductor (SNS) junctions, which was originally analyzed in Ref.\,\cite{PhysRevB.98.075425}. In contrast to the NS junctions discussed above, in this case the superconducting phase $\phi$ cannot be gauged away. In fact, if there is a finite phase-difference between the two superconductors, ARs at both interfaces lead to the formation of Andreev bound states (ABSs) within the gap \cite{kuliksns}. The number of ABSs depends on the ratio between the coherence length $\xi$ and the length of the N region $L_{\rm N}$. For $L_{\rm N}\ll\xi$ a pair of ABSs emerge, while for $L_{\rm N}\gg\xi$ the junction hosts many levels. In Ref.\,\cite{PhysRevB.98.075425}, short junctions ($L_{\rm N}\ll\xi$) were studied with the left (L) and right (R) S regions assumed to be conventional spin-singlet $s$-wave superconductors with $\Delta_{\rm L(R)}=\Delta\,{\rm e}^{i\phi_{\rm L(R)}}$, where $\phi_{\rm L(R)}$ is the superconducting phase. Without loss of generality, a finite phase difference $\phi$ across the junction is generated by setting $\phi_{\rm L}=0$ and  $\phi_{\rm R}=\phi$. The pair amplitudes in this case were calculated following a scattering approach, similar to the one discussed for NS junctions in the previous subsection. The scattering states were defined for each region, matched at the interfaces, and then the retarded Green's function were obtained using the appropriate boundary conditions. 

As shown in Ref.\,\cite{PhysRevB.98.075425}, the pair amplitudes appearing in SNS junctions share some similarities with the pair amplitudes in the S region of NS junctions. As one might expect, the induced pair amplitudes in the left and right S regions of short SNS junctions possess contributions from both the bulk and the interface, as we discussed for the S region of NS junctions, with the interface terms emerging due to NRs and ARs. Also, the role of ARs is similar to the situation in NS junctions in that they allow for the coexistence of ESE and OSO pair amplitudes, while NRs only contribute to the ESE pairing. 

Apart from these similarities to NS junctions, the pair amplitudes in short SNS junctions also display some crucial differences. Importantly, the NR and AR scattering processes acquire a non-trivial dependence on the phase difference across the junction, $\phi$, which they impart to the interface pair amplitudes. Moreover, since it is well-known that ARs and NRs capture the emergence of ABSs \cite{kuliksns,Beenakker:92}, the pair amplitudes also indirectly reflect the presence of ABSs, as demonstrated in Ref.\,\cite{PhysRevB.98.075425}. The exact calculations in Ref.\,\cite{PhysRevB.98.075425} show that the ESE and OSO interface amplitudes exhibit different phase-dependent interference profiles in their AR contributions which can be easily distinguished. Furthermore, a finite supercurrent flows across the short SNS junction when $\phi\neq0$ and its value is determined by the ABSs \cite{PhysRevLett.67.132,FURUSAKI1991299}. Based on this, it is possible to correlate the phase-dependent profile of the supercurrent with the specific phase-dependent interference features of the ESE and OSO pairings\cite{PhysRevB.98.075425}. 

We finally note that, within this above described scattering approach, the pair amplitudes represent intralead correlations, existing in either the left or right S region, and, hence, the possibility of interlead correlations is not considered. Recently it was shown that, in addition to the intralead odd-$\omega$ pairing discussed above for short SNS junctions, interlead odd-$\omega$ pairing can emerge in tunneling junctions with conventional superconductors \cite{Triola18}. In this case, the left and right superconductors introduce an additional index to the classification of Cooper pair symmetries. The authors showed that this interlead odd-$\omega$ pairing is linear in the interlead tunneling, while traditional intralead pairing is quadratic. Therefore, at small tunneling amplitudes, the odd-$\omega$ interlead pairing becomes the dominant channel near the junction whenever there is a finite Josephson current.

\section{Odd-$\omega$ pairing in nanowires with Rashba spin-orbit coupling}
\label{sec3}
In this section we discuss the possibility of inducing spin-triplet correlations in 1D junctions by adding Rashba SOC. Rashba SOC arises due to a lack of structural inversion symmetry in some materials \cite{PhysRev.100.580,Rashba1960,rashba84a} and induces a non-trivial coupling between the crystal momentum and the electronic spin degrees of freedom. This spin-momentum mixing can drive the conversion of spin-singlet to spin-triplet pair amplitudes \cite{PhysRevLett.87.037004,PhysRevLett.92.027003,PhysRevLett.92.097001,Reyren07,doi:10.1143/JPSJ.76.051008,PhysRevB.79.094504,PhysRevLett.113.227002,0034-4885-80-3-036501}. Hence, the combination of SOC and conventional spin-singlet superconductors is expected to favor the emergence of spin-triplet pair correlations without the need for magnetism.

Although SOC can be controlled by voltage gates \cite{0268-1242-11-8-009,PhysRevLett.78.1335,doi:10.1021/nl301325h,Takase17}, the main motivation to study junctions based on nanowires with Rashba SOC comes from recent experiments which have shown that large intrinsic Rashba SOC is present in InAs \cite{chang15,Higginbotham,Krogstrup15,Deng16,Albrecht16,vaitienkenas17,deng18}, InSb \cite{zhang16,0957-4484-26-21-215202,Gazibegovic17,zhang18}, and InAsSb nanowires \cite{PhysRevMaterials.2.044202}.
Moreover, these experiments on nanowires have shown a strong proximity effect, characterized by hard induced gaps, thus providing the required superconducting order parameter in the nanowire and making these systems key candidates for topological superconductivity. Inspired by these advances, it has recently been shown that odd-$\omega$ spin-triplet pairing emerges in 1D nanowire NS and short SNS junctions with Rashba SOC formed by proximity to conventional $s$-wave spin-singlet superconductors \cite{PhysRevB.92.134512,PhysRevB.98.075425}. 

\begin{figure}
  \begin{center}
\includegraphics[width=.85\textwidth]{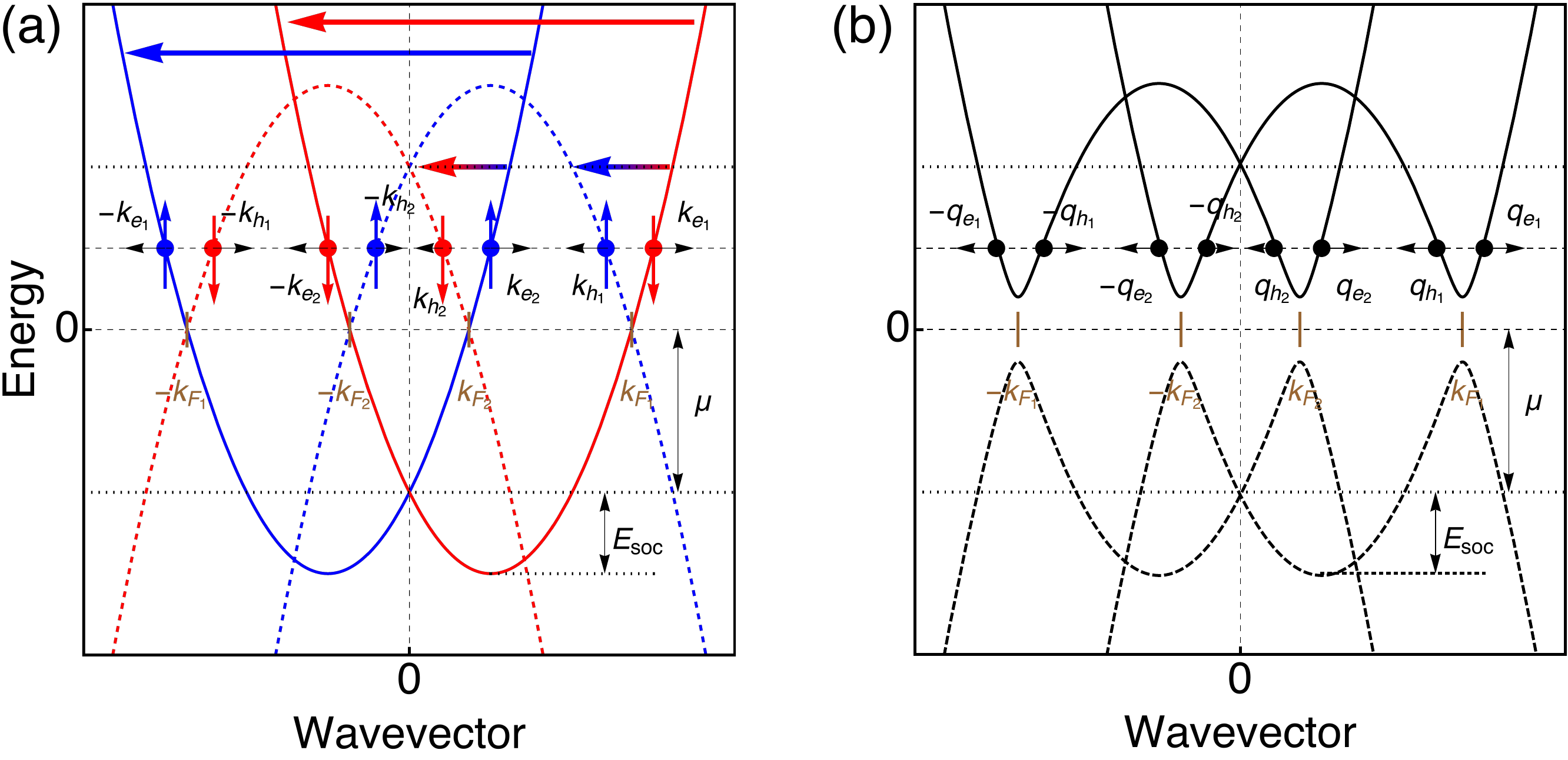} 
  \end{center}
\caption{(a)  Energy versus momentum dispersion in the normal part (N) at $\Delta=0$, where the Rashba SOC splits the normal bands around zero momentum $k=0$. Solid (dashed) blue and red parabolas correspond to electrons (holes) with spin up and down, respectively.  The bottom (top) of the electron (hole) band is indicated by dotted horizontal lines.
(b) A finite order parameter $\Delta \neq 0$ in the superconductor (S) opens a gap at the Fermi momenta $\pm k_{\rm F_{1(2)}}$ and  mix electron and hole bands with different spins, leading to quasiparticle bands with positive (solid curves) and negative (dashed) energies. At the NS interface eight  scattering processes from incident particles towards the interface occur, four coming from N and four from S, which enable the  calculation of the Green's function.
This can be visualized by fixing the energy to a positive value, indicated by  horizontal dashed line above zero energy, giving rise to eight values of momentum, $\pm k_{e_{1,2}}$ and $\pm k_{h_{1,2}}$ in N and $\pm q_{e_{1,2}}$ and $\pm q_{h_{1,2}}$ in S, where small horizontal black arrows denote the direction of motion of the  states associated to each momentum. The four processes from N account for right moving  particles (spin up and down electrons, spin up and down holes) towards the interface. Likewise, the four processes from S correspond to left moving particles towards the interface. The scattering processes are similar to the ones described in Fig.\,\ref{NS} with the difference that here spin needs to be taken into account. For instance, NR takes place between states of the same spin band, with different momenta e.g.~$k_{\rm e_{1(2)}} \rightarrow -k_{\rm e_{2(1)}}$, indicated by blue and red horizontal arrows, while AR occurs between states of different spin bands with momenta, e.g.~$k_{\rm e_{1(2)}} \rightarrow k_{\rm h_{1(2)}}$, shown by blue-red and red-blue arrows. Reprinted modified figure with permission from [J. Cayao and A. M. Black-Schaffer, Phys. Rev. B 98, 075425 (2018)] Copyright (2018) by The American Physical Society. 
}
\label{figrashba}
   \end{figure}

Assuming that the 1D nanowire with Rashba SOC possesses only a single-mode, it can be modelled effectively, in the basis $(\psi_{\uparrow},\psi_{\downarrow},\psi_{\uparrow}^{\dagger},\psi_{\downarrow}^{\dagger})$, using the BdG Hamiltonian: 
\begin{equation}
\label{HBdGRashba}
H_{\rm 1DRashba}=\Big(\frac{p_{x}^{2}}{2m}-\mu_{i}\Big)\sigma_{0}\tau_{z}+\frac{\alpha}{\hbar}\sigma_{z}\tau_{0}p_{x}+\Delta(x)\sigma_{y}\tau_{y}\,,
\end{equation}
where $p_{x}=-i\hbar\partial_{x}$, while $\sigma_{i}$ and $\tau_{i}$ are the $i$-Pauli matrices in spin and Nambu spaces, respectively,  $\alpha$ is the Rashba SOC strength, and $\Delta(x)$ the order parameter with finite value in S while zero in N. The SOC splits the normal spin bands around $k=0$, while a finite $\Delta$ opens gaps at the Fermi momenta $k_{F_{1,2}}\approx k_{\rm SOC}+\bar{k}$, $\bar{k}=\sqrt{2m(\mu_{i}+E_{\rm SOC})/\hbar^{2}}$, mixing electron and hole bands of different spins as seen in Fig.\,\ref{figrashba}. Here $E_{\rm SOC}=m\alpha^{2}/(2\hbar^{2})$ is the SOC energy. 

Following the scattering approach introduced in Section~\ref{sec2},  the pair correlations at finite SOC can be obtained. 
Again, we stress that this approach is both conceptually simple and experimentally relevant as it utilizes scattering processes, such as NR and AR, that can be obtained in conductance measurements \cite{chang15,doi:10.1063/1.4971394,gulonder,Gazibegovic17,zhang18}. There are eight scattering processes in this case due to incident particles, at the NS interface, which correspond to four states in N with $k_{\rm e_{1,2}}$ and $-k_{\rm h_{1,2}}$ and also to four states in S with $-q_{\rm e_{1,2}}$ and $q_{\rm h_{1,2}}$, as shown in Fig.\,\ref{figrashba}. 
The problem is analytically solvable but the expressions for the pair amplitudes are quite cumbersome, we therefore here only highlight the main effects of SOC and refer the reader to \cite{PhysRevB.98.075425} for the full analytic expressions and details of the calculations. 

\begin{figure}
  \begin{center}
\includegraphics[width=.75\textwidth]{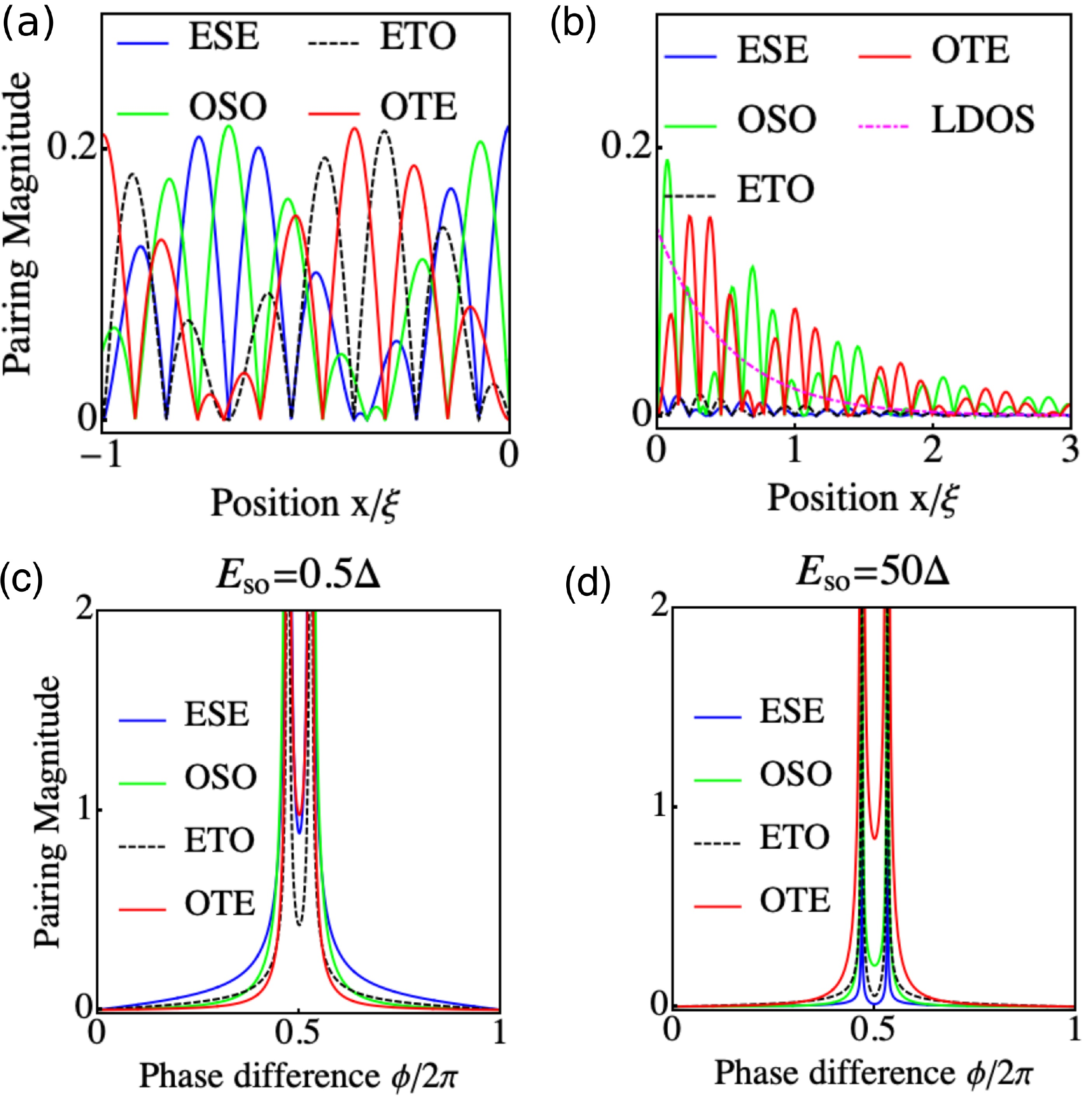} 
  \end{center}
\caption{(a,b) Induced interface pair magnitudes as function of distance from the interface in the N (a) and  S (b) regions of NS junctions with Rashba SOC for $x'=0$, $\omega=0.1\Delta$, $E_{\rm SOC}=0.5\Delta$, and $\mu_{N,S}=10\Delta$. Magenta curve in (b) also shows the interface LDOS. 
(c,d) Induced interface pair magnitudes in short SNS junctions as function of phase difference at weak (c) and strong Rashba SOC (d) for $x=0.1\xi$,  $x'=0$, $\omega=0.1\Delta$, and $\mu_{\rm N, S}=10\Delta$. 
Reprinted figure with permission from [J. Cayao and A. M. Black-Schaffer, Phys. Rev. B 98, 075425 (2018)] Copyright (2018) by The American Physical Society. 
}
\label{figrashba1}
   \end{figure}

We begin our discussion of the pair amplitudes by noting that the spin-singlet components (ESE and OSO) arise due to the mixing of spatial parities at the NS interface, in agreement with our discussion in Section \ref{sec2}. Interestingly, due to the spin-mixing caused by the SOC, spin-triplet correlations (ETO, OTE) also emerge at interfaces. All these symmetry classes (ESE, OSO, ETO, OTE) correspond to the pairings allowed by Fermi-Dirac statistics, as shown in Table \ref{tab1}, and they coexist at the NS interface. This is illustrated in Fig.\,\ref{figrashba1} (a,b), displaying the pair magnitudes in N and S, respectively, due to the interface (i.e.~not bulk contributions).  However, locally in space, i.e.~at $x=x'$, the features specific to the 1D Rashba SOC nanowire induce vanishing spin-triplet correlations, even the interface OTE pairing despite being even parity \cite{PhysRevLett.113.227002,PhysRevB.92.134512,PhysRevB.98.075425}. 

In the N region the pair amplitudes are solely determined by ARs, while in the S region the amplitudes have both bulk and interface contributions, with the interface terms coming from both NRs and ARs. It is here appropriate to also mention that the Rashba SOC induces ETO pairing which is not proportional to any scattering process occurring at the interface, but is instead of bulk nature, as also established in Refs.\,\cite{PhysRevLett.87.037004,PhysRevB.81.184502}. 
Analyzing the scattering contributions further, NRs only contribute to the even-$\omega$ pairing in S, while AR mediates the emergence of all pair symmetries in both N and S. This can be understood from the processes involving two reflections at NS interfaces, namely, the NR and AR indicated by blue/red and red-blue/blue-red gradient horizontal arrows in Fig.\,\ref{figrashba}. In fact, AR takes place between states of different spin bands with momenta $k_{e_{1(2)}}\rightarrow k_{h_{1(2)}}$. This leads to AR contributions in the pair amplitudes that mix spatial coordinates ($x$ and $x'$) with electron and hole wave vectors of different spin bands ($-k_{\rm e_{1(2)}}$ and $k_{\rm h_{1(2)}}$), facilitating the generation of even- and odd-frequency components. The minus sign in $-k_{\rm e_{1(2)}}$ comes from the different direction of motion of electrons and holes during the AR process. The role of AR is thus the same as discussed in Section \ref{sec2} without Rashba SOC. A similar analysis applies for the NR, but taking into account the fact that it occurs between states of the same spin band with different momenta e.g.~$k_{e_{1(2)}}\rightarrow -k_{e_{2(1)}}$, implying that NR contributions emerge only with even parity.
Here, $k_{e_{1(2)}}(\omega)=k_{\rm F_{1(2)}}+\kappa^{\rm N}$ and $k_{h_{1(2)}}(\omega)=k_{e_{1(2)}}(-\omega)$, with $\kappa^{\rm N}=\bar{k}\omega/[2(\mu_{i}+E_{\rm SOC})]$, $k_{SOC}=m\alpha/\hbar^{2}$.

In terms of spatial properties, the pair amplitudes in N exhibit an oscillatory profile in space, while in S the interface pair amplitudes also oscillate in space but with an exponentially decaying amplitude, as seen in Fig.\,\ref{figrashba1} (a,b). 
The short-period oscillations are due to the chemical potential, as discussed in Section \ref{sec2} for zero SOC, but there exist also long-period oscillations due to the SOC. This SOC effect leads to a visible beating feature in the pairing magnitudes. The amplitudes in N do not decay due to the zero temperature assumption considered in the scattering method, as explained in Sec.\,\ref{sec2}.
Furthermore, at low frequencies the OSO and OTE amplitudes in fact acquire larger values than the even-$\omega$ terms mainly due to the large contribution from the AR. In this regime the interface LDOS also has a strong contribution from AR and exhibits an exponential decay from the interface but without oscillations due to negligible NRs, as shown by the magenta curve in Fig.\,\ref{figrashba1}(b). 

Next, we briefly comment on transparent short SNS junctions. The interface pair amplitudes vanish at $\phi=0$ \cite{PhysRevB.98.075425}. An interesting regime appears for strong SOC at $\phi=\pi$, where even-$\omega$ pairing is much smaller than the  odd-$\omega$ pairing for frequencies between the energies of the ABS and the gap $\Delta$, as seen in Fig.\,\ref{figrashba1}(c,d) for  a specific frequency within the said regime. This occurs because, at strong SOC, NRs are reduced and, although ARs still give contributions of the same magnitude, their contributions interfere destructively for even-$\omega$ and constructively for odd-$\omega$-pairing, leading to larger odd-$\omega$ amplitudes. 

To conclude this section, we draw further attention to the fact that odd-$\omega$ mixed-spin triplet pairing emerges due to the Rashba SOC in junctions without any magnetism. Its potential for characterization based on the observation of scattering processes, namely, NR and AR, is particularly relevant as their amplitudes can be obtained from conductance measurements in the recently realized 1D Rashba nanowires \cite{chang15,doi:10.1063/1.4971394,gulonder,Gazibegovic17,zhang18,Aguadoreview17,LutchynReview08,zhangreview}. 
By including a magnetic region or magnetic field into the junctions, topological superconductivity can also be induced and equal-spin spin triplet correlations appear. However, we leave this discussion to Section \ref{sec5} where we in detail address the relation between odd-$\omega$ pairing and topological superconductivity.

\section{Odd-$\omega$ pairing at the edges of 2D topological insulators}
\label{sec4}
In this section we discuss the possibility of generating odd-$\omega$ spin-triplet correlations using the 1D edges of 2D TIs, which are characterized by the presence of an insulating band gap throughout the 2D bulk but with metallic states along their 1D perimeter. An important feature of these 1D edge states is that they possess perfect helicity such that counter-propagating edge modes carry opposite spin and, hence, backscattering is forbidden \cite{PhysRevLett.95.146802,PhysRevLett.95.226801,Bernevig06,PhysRevLett.96.106802}.
These states have already been experimentally realized in both HgCd/HgTe \cite{konig07,Roth09,brune2012,Nowack2013} and InAs/GaSb \cite{PhysRevLett.100.236601,PhysRevLett.107.136603} heterostructures. Here we focus on superconducting properties of the 1D edges, for general reviews on topological insulating states see \cite{RevModPhys.82.3045,RevModPhys.83.1057,doi:10.1002/pssb.201248385,Ando13,Sato_2017,tkachov19review}. 

By combining 2D TIs in junctions with conventional $s$-wave superconductors novel pair correlations can be induced by proximity in the metallic edge states \cite{PhysRevLett.100.096407,PhysRevB.79.161408}, giving rise to a topological superconducting phase, as also demonstrated experimentally \cite{Yacoby14,vlad15}. Importantly, some of the novel pair correlations in these systems have been predicted to be odd-$\omega$ mixed spin-triplet 
\cite{PhysRevB.86.075410,PhysRevB.86.144506,PhysRevB.87.220506,PhysRevB.92.100507,PhysRevB.96.155426,PhysRevB.96.174509,PhysRevB.97.075408,PhysRevB.97.134523}. These results helped to understand proximity-induced superconductivity in TIs  and showed that odd-$\omega$ correlations should be included in the study of  low energy states and MZMs. We now illustrate how odd-$\omega$ pairing can emerge in 2DTIs to highlight its fundamental importance for understanding topological states.

\begin{figure}
  \begin{center}
\includegraphics[width=.85\textwidth]{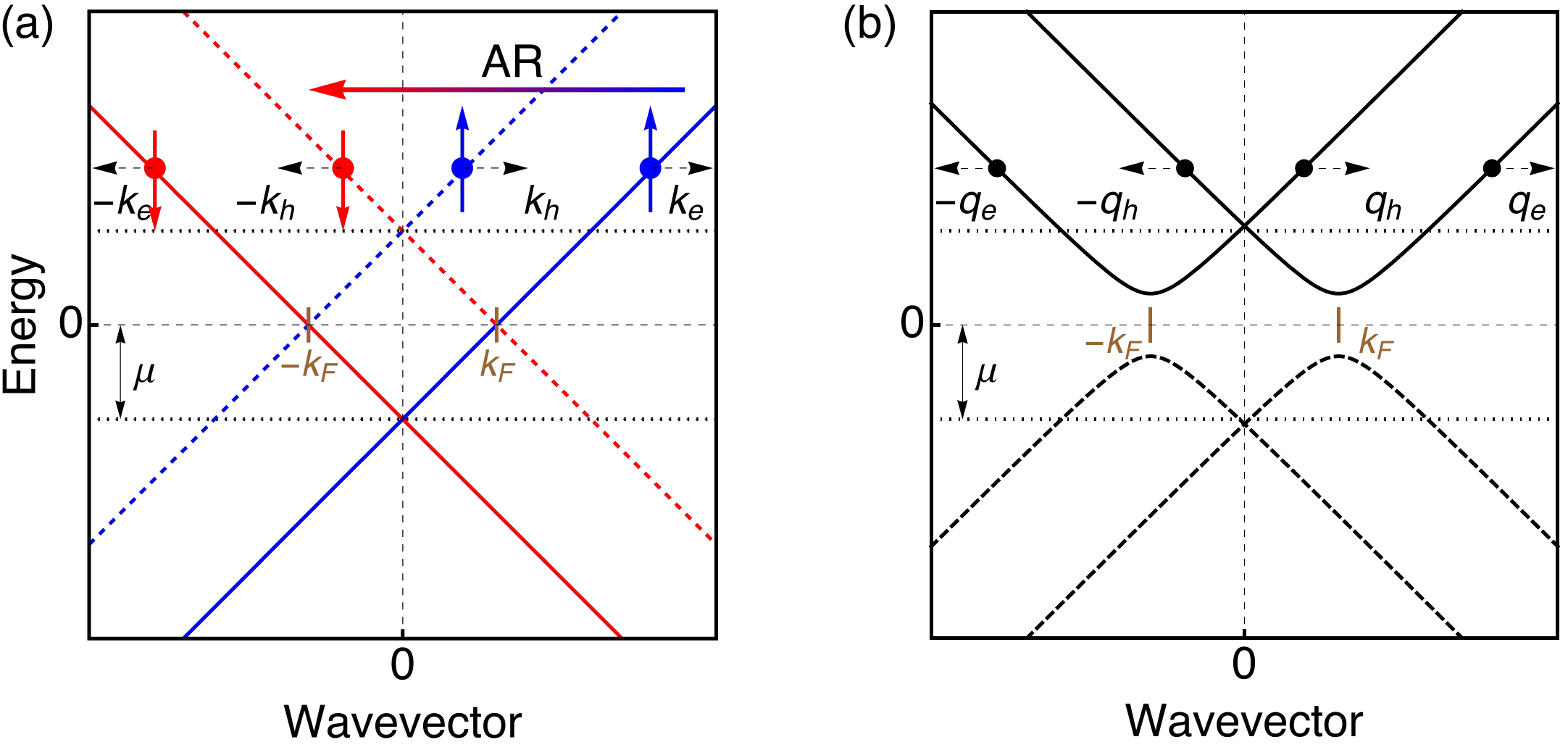} 
  \end{center}
\caption{(a) Energy versus momentum dispersion for the 1D metallic edge of a 2D TI in the normal part (N) at $\Delta=0$, where solid (dashed) red and blue curves
 corresponds to electrons (holes). The chemical potential for electron (hole) band  is indicated by dotted horizontal lines. (b) A finite order parameter $\Delta \neq 0$ in the superconductor (S) opens a gap at the Fermi momenta $\pm k_{\rm F}$ and mixes electrons and hole bands with different spins. Solid (dashed) black  curves denote quasiparticle bands with positive (negative) energies.  At the NS interface four scattering processes from incident particles towards the interface take place, two from N and two from S. This can be visualized by fixing the energy to a positive value, indicated by horizontal dashed line above zero energy, giving rise to four values of momentum $\pm k_{e, h}$ in N and four $\pm q_{e,h}$ in S, where small horizontal black arrows in N and S denote the direction of motion of the states associated with each momentum.
These processes are then used to calculate the Green's function as discussed in previous two sections. Due to the helical nature of the metallic edge, where particles with spin-up move along one direction and particles with spin-down along the other, NR are forbidden and only AR occurs as indicated by the blue-red solid arrow.
Reprinted modified figure with permission from [J. Cayao and A. M. Black-Schaffer, Phys. Rev. B 96, 155426 (2017)] Copyright (2017) by The American Physical Society. 
}
\label{fig2DTI0}
   \end{figure}

The emergence of odd-$\omega$ correlations in 2D TIs can be understood based on a simple low-energy effective Hamiltonian for the 1D metallic edges with spin quantization along the $z$-axis:  
\begin{equation}
\label{2dti}
H_{\rm 1Dedge}= v_{F}p_{x}\tau_{0}\sigma_{z}-\mu\tau_{z}\sigma_{0}+ \Delta(x)\tau_{y}\sigma_{y}\,,
\end{equation} 
expressed in the basis $(\psi_{\uparrow},\psi_{\downarrow},\psi_{\uparrow}^{\dagger},\psi_{\downarrow}^{\dagger})$ and where $\sigma_{i}$ and $\tau_{i}$ are the $i$th Pauli matrices in spin and Nambu spaces, respectively, and $\Delta$ is the order parameter induced by proximity to a superconducting region. In what follows, the order parameter $\Delta(x)$ is assumed to vanish when the helical edge is coupled to a normal region N and is assumed to take on a finite value when coupled to a superconducting region S. We here discuss cases where an NS and short SNS junction are formed along the 1D edge.

The helical nature of the metallic edge states can be seen from the energy-momentum dispersion of Eq.\,(\ref{2dti}) and is also shown in Fig.\,\ref{fig2DTI0}(a,b). 
In the normal state ($\Delta = 0$)  the energy levels of these edges disperse linearly with momentum and the direction of the spin (up/down) is locked with the direction of the momentum (right/left), as seen in Fig.\,\ref{fig2DTI0}(a). When these edges are coupled to a superconducting region ($\Delta\neq 0$) a gap opens at the Fermi momenta mixing the electron and hole branches with different spins, as depicted in Fig.\,\ref{fig2DTI0}(b). Moreover, it has been theoretically and experimentally found that a consequence of helicity in NS junctions at the metallic edge is that NRs (or equivalently backscattering) are forbidden and only perfect (local) AR occurs for energies within the superconducting gap \cite{PhysRevB.79.161408,PhysRevB.82.081303,PhysRevLett.109.186603}. 
Before proceeding further, we observe that the second term in Eq.\,(\ref{HBdGRashba}) and the first term in Eq.\,(\ref{2dti}) have the same structure and linear momentum dependence. The linear momentum term in Eq.\, (\ref{HBdGRashba}) was crucial for the generation of spin-triplet correlations in junctions with Rashba SOC discussed in the previous section, and therefore naturally, spin-triplet correlations are also expected for the 1D helical edge in Eq.~(\ref{2dti}). 

The pair amplitudes for NS and SNS junctions for Eq.~(\ref{2dti}) have been calculated based on the scattering Green's function approach in Ref.\,\cite{PhysRevB.96.155426}, as discussed in previous sections. We here focus on their main findings and refer to the original work for explicit analytical expressions and details of the calculations.
\begin{figure}
  \begin{center}
\includegraphics[width=.45\textwidth]{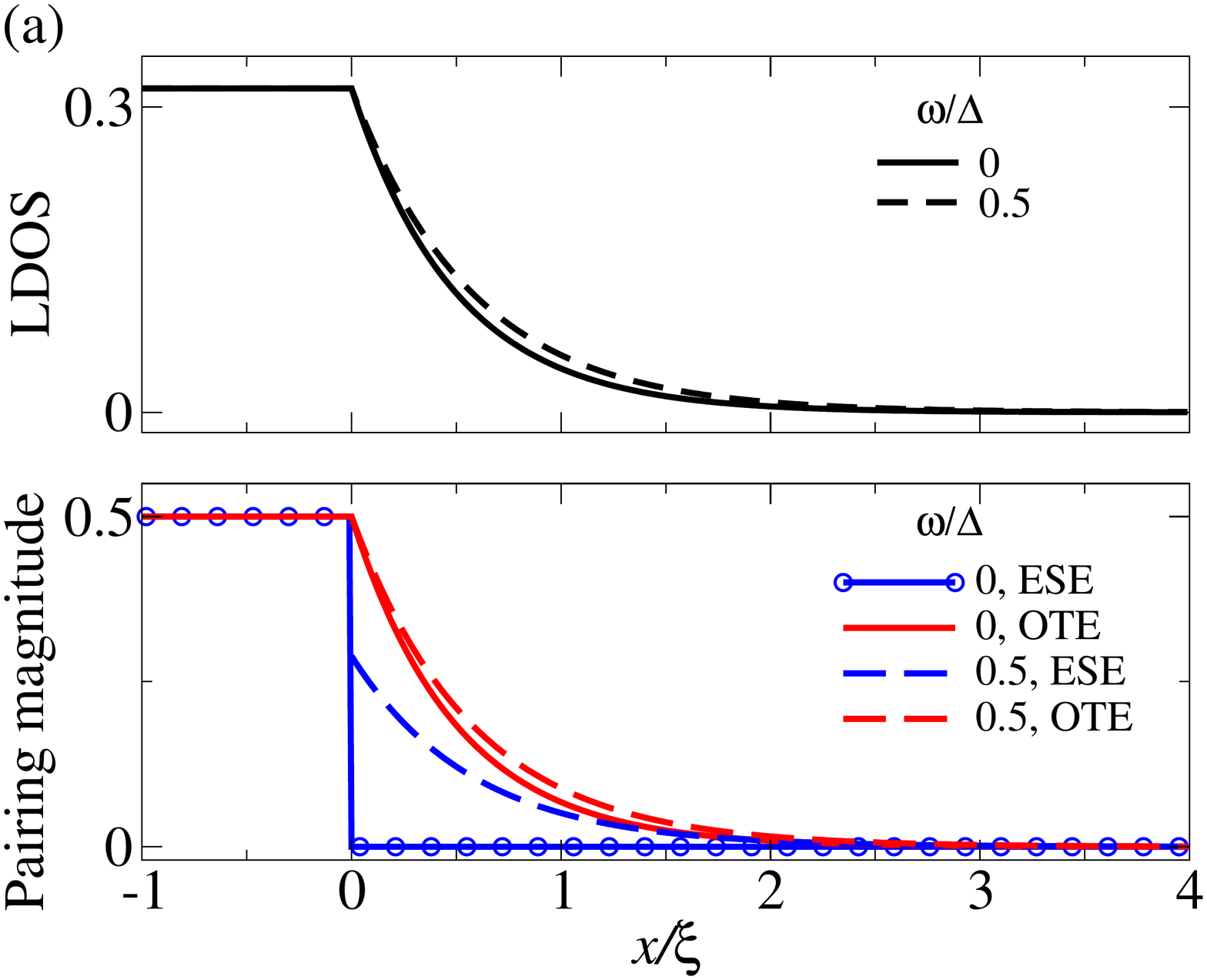} 
\includegraphics[width=.45\textwidth]{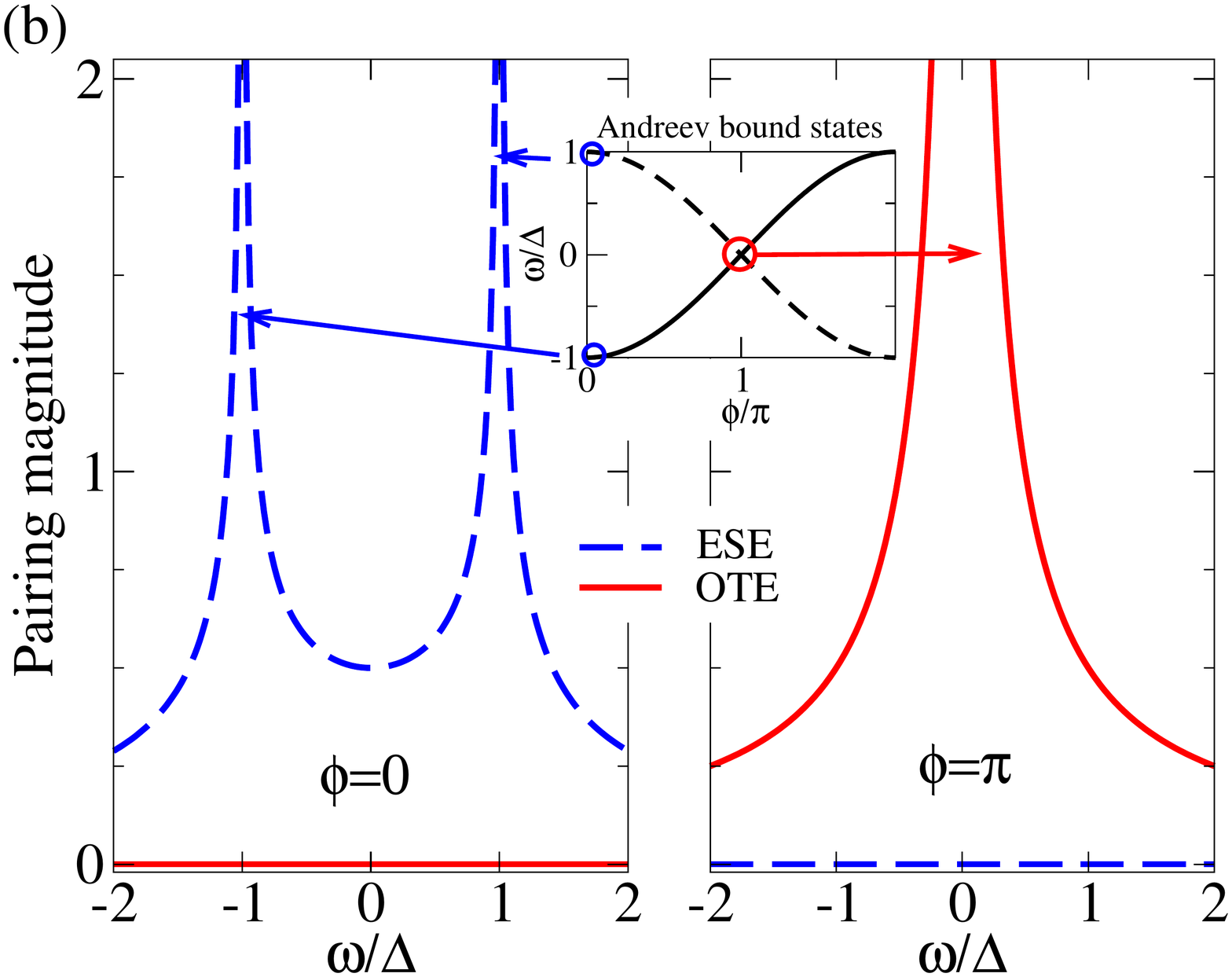} 
  \end{center}
\caption{(a) Spatial dependence of the LDOS  (top) and even-$\omega$ spin-singlet (ESE) and odd-$\omega$ spin-triplet (OTE) $s$-wave (local) pair magnitudes in a NS junction (bottom). In both panels, solid curves correspond to $\omega=0$ while dashed curves to $\omega=0.5\Delta$.
 (b) Frequency dependence of ESE and OTE pair magnitudes in the normal region of a very short SNS junction at $\phi=0$ (left) and $\phi=\pi$ (right). Inset: Lowest ABSs as a function of $\phi$. Reprinted figure with permission from [J. Cayao and A. M. Black-Schaffer, Phys. Rev. B 96, 155426 (2017)] Copyright (2017) by The American Physical Society. 
}
\label{fig2DTI}
   \end{figure}

For an NS junction located at $x=0$ there is a coexistence, at the interface, of pair amplitudes in all symmetry classes in Table \ref{tab1}, i.e.~ESE, OSO, ETO, and OTE. 
In this case the ESE and OSO amplitudes emerge due to spatial symmetry breaking as in NS junctions with normal metals discussed in Section\,\ref{sec2}. 
The spin-triplet correlations (ETO, OTE) correspond to mixed spin-triplet, as in the case of Rashba nanowires, but here they emerge due to the helicity of the 1D metallic edge of 2D TIs. The metallic edge thus allows for mixing different spin states without magnetism.  
Since NRs are forbidden, due to helicity, ARs are the sole contributors to this coexistence of all pairing classes at the interface. 

In terms of the pair amplitudes in the S region, bulk terms with both ETO and ESE symmetries exists, as in junctions with Rashba SOC discussed in the previous Section. The emergence of the ETO correlations in the bulk is also consistent with Refs.\,\cite{PhysRevB.81.184502,PhysRevB.81.241310,PhysRevB.88.075401,PhysRevB.92.205424}. Additionally, near the interface (I) within the S region, the pair amplitudes acquire contributions from all the symmetry classes (ESE, OSO, ETO, OTE) due, entirely, to AR processes.
Interestingly, locally in space $x = x'$, only the interface pairings with ESE and OTE symmetry are finite and given by $f_{0, {\rm I}}^{r,{\rm E}}(x,\omega)=[-r_{eh}(\omega)/(2iv_{\rm F})]{\rm e}^{-2\kappa(\omega)x}B(\omega)$ and $f_{3, {\rm I}}^{r,{\rm O}}(x,\omega)=[-r_{eh}(\omega)/(2iv_{\rm F})]{\rm e}^{-2\kappa(\omega)x}$. Here, $r_{eh}={\rm e}^{-i\eta(\omega)}$ is the AR coefficient, $\eta(\omega)={\rm arcos}(\omega/\Delta)$, $v_{\rm F}$ is the Fermi velocity,  $\kappa(\omega)=\sqrt{\Delta^{2}-\omega^{2}}/v_{\rm F}$, and $B(\omega)$ is a frequency dependent coefficient. The spatial behavior of these terms is shown in the bottom panel of Fig.\,\ref{fig2DTI}(a). They are finite due to the strong spin-momentum locking of 2DTIs, where counter-propagating edge modes carry opposite spin. This is in stark contrast to Rashba nanowires where all such local spin-triplet terms vanish. Moreover, at low frequencies, the OTE amplitude is clearly the dominant pairing channel. 
We also note that in the helical edge the interface pair amplitudes within S decay exponentially into the bulk but without any oscillatory behavior, unlike the junctions with Rashba nanowires. 

The LDOS was also investigated in Ref.~\cite{PhysRevB.96.155426} for this setup which, for $|\omega|<\Delta$, has the form $\rho(x,\omega)(1/\pi v_{\rm F}){\rm e}^{-2\kappa(\omega)x}$ in the S region, shown in the top panel of Fig.\,\ref{fig2DTI}(a).
For the same range of frequencies, $|\omega|<\Delta$, the AR coefficient is maximized $r_{eh}=1$, and, therefore, by simple inspection we notice that the OTE amplitude and LDOS exhibit the same spatial dependence, suggesting a strong connection between these two quantities, as also seen by the enhancement close to the interface in S in Fig.\,\ref{fig2DTI} (a). 
We here also emphasize that the AR coefficients can be obtained directly from the conductance $G=2e^{2}/h(1+ |r_{eh}|^{2}$). Hence, these two well-established \cite{PhysRevLett.109.186603,Yacoby14,vlad15,Bocquillon17} experimental observables, LDOS and conductance, allow for a simple determination of the pair amplitudes, which can otherwise be notoriously hard to obtain.

We next turn our attention to the case of a helical edge coupled to a short SNS junction. In this setup, all pairing classes discussed above are similarly induced but now with a strong dependence on the phase difference across the junction, $\phi$, which also relates the pair amplitudes to the emergence of ABSs \cite{PhysRevB.96.155426,PhysRevB.97.075408}.
This can be understood from the fact that ARs are the only processes that determine the induced pairing at the interfaces and they also reflect the emergence of ABSs \cite{PhysRevLett.100.096407,PhysRevB.86.214515}.  
This relationship is illustrated in Fig.\,\ref{fig2DTI}(b), where ESE and OTE magnitudes are plotted in the N regime in short junctions. In short SNS junctions, the ABSs energy levels are given by $\omega_{\pm}(\phi)=\pm \Delta{\rm cos}(\phi/2)$ and exhibit a $4\pi$ periodicity due to the zero-energy crossing at $\phi=\pi$, which is protected by time-reversal symmetry \cite{PhysRevLett.100.096407,PhysRevB.79.161408,PhysRevB.86.214515,PSSB:PSSB201248385,PhysRevB.88.075401}. The zero-energy ABS is its own charge-conjugate state and corresponds to a MZM which is twofold degenerate. Interestingly, the special features of ABSs are captured in the form of resonances in the pair magnitudes, as shown in Fig.\,\ref{fig2DTI}(b).
Indeed, at $\phi=0$, the OTE amplitude is zero and the ESE reveals the gap edges which merge with the continuum.  On the other hand, at $\phi=\pi$, ESE pairing vanishes and OTE pairing exhibits a resonance peak at $\omega=0$. The vanishing of the OTE (ESE) amplitude at $\phi=0(\pi)$ results from destructive interference processes in the N region. Remarkably, the conditions for the resonance peak in OTE pairing at $\phi = \pi$ correspond exactly to the protected zero-energy crossing of the ABSs \cite{PhysRevB.96.155426,PhysRevB.97.075408}. 

While the above discussions of odd-$\omega$ pairing were performed using a low-energy continuum model to describe the helical edges of a 2D TI coupled to a superconductor, similar results have also been found using lattice models. Specifically, in Ref.~\cite{PhysRevB.96.174509} the emergence of odd-$\omega$ pairing was explored using a lattice model describing a buckled 2D honeycomb lattice proximity-coupled to a conventional $s$-wave superconductor. In that work, the authors studied both the topologically trivial regime in which bulk 2D odd-$\omega$ pairing can be realized, as well as the low-doping regime in which the low-energy states are localized along the edges. In the latter regime, odd-$\omega$ pairing was actually found to emerge generically at the helical edge even in the absence of a NS or SNS junction along the edge. Instead, a finite gradient in $\Delta$ emerges naturally at the edges of buckled honeycomb lattices due to the sublattice structure at the edge termination, in agreement with earlier work showing how odd-$\omega$ pairing appears for finite gradients in TI surface and edge states  \cite{PhysRevB.86.144506}. 

\begin{figure}
  \begin{center}
\includegraphics[width=.95\textwidth]{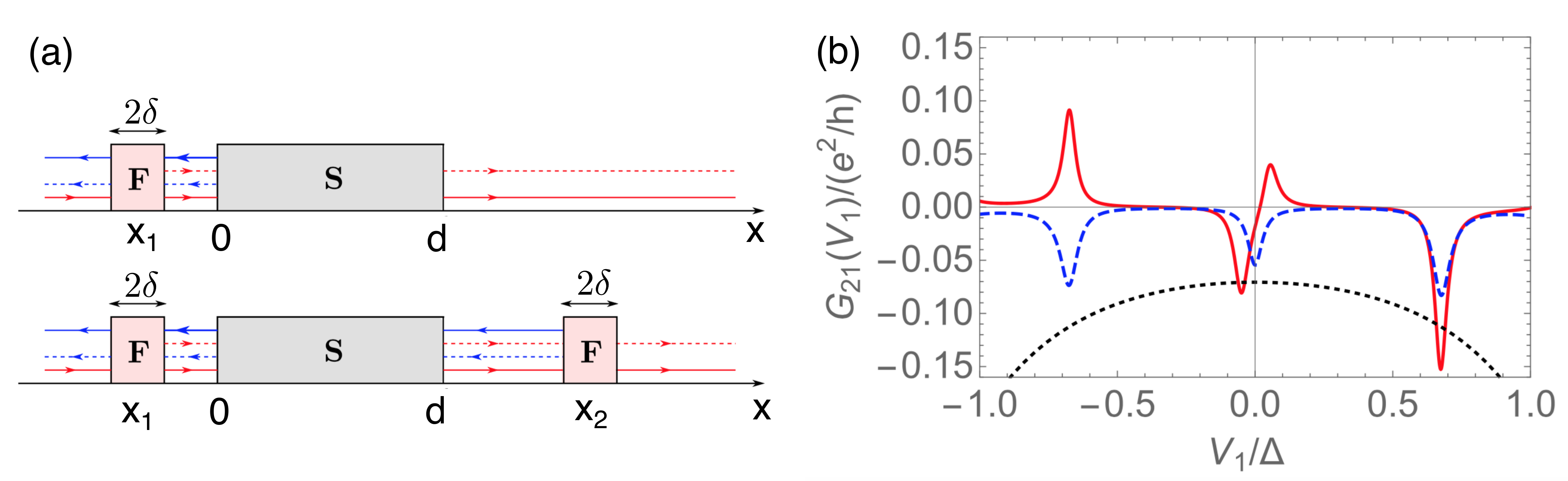} 
  \end{center}
\caption{(a) FS and FSF junctions at the helical edge of a 2D TI, with the F regions are located at $x_{1}$ and $x_{1,2}$, respectively. The length of F is $2\delta$, while the S region have length $d$. (b) Non-local conductance for FS (blue) and  FSF (red) as a function of the bias between F and S, $V_{1}$.  A positive signal in the non-local conductance indicates an excess of CAR over ET. Black dotted line shows the conductance in the absence of any ferromagnetic impurity. Reprinted figure with permission from [F. Cr\'{e}pin, P. Burset, and B. Trauzettel, Phys. Rev. B 92, 100507(R) (2015)] Copyright (2015) by The American Physical Society.  
}
\label{crepin}
   \end{figure}
   
\subsection{Equal-spin triplet pairing}
So far, in this section we have focused on the emergence of odd-$\omega$ pairing due to the helicity of the metallic edges of 2D TIs but explicitly in the absence of magnetism. However, a crucial requirement for superconducting spintronics are equal-spin triplet, or spin-polarized, pair correlations, which have largely been discussed in context of FS junctions \cite{LinderNat15,7870d3ff91ed485fa3e55e901ff81c80,EschrigNat15,0034-4885-78-10-104501}. 
When the edge of a 2D TI is also in proximity to a ferromagnet (F) as depicted in Fig.\,\ref{crepin}(a), NR and crossed AR (CAR), which would be otherwise suppressed by helicity, are allowed.
The authors of Refs.\,\cite{PhysRevB.92.100507,PhysRevB.97.075408} have demonstrated that, in FS junctions equal-spin triplet OTE pairing can appear as a result of equal-spin CAR, in addition to the spin-singlet and mixed spin-triplet correlations already discussed. 
This occurs because, in a CAR process, a right moving electron from the left lead can be converted into a right moving hole with the same spin in the right lead, see Fig.\,\ref{crepin}(a), which  corresponds to a Cooper pair being injected in the equal-spin channel \cite{PhysRevB.74.180503,PhysRevB.90.205435}. As a consequence, CAR processes directly probe the pairing induced by the proximity of F. 
CAR can be measured in the non-local conductance, which, in general, is given by $G_{21}=(e^{2}/h)(CAR-ET)$, where ET denotes electron transmission processes which might mask the effect of CAR. The authors of Ref.\,\cite{PhysRevB.92.100507} showed that although CAR and ET compete in the non-local conductance, in FSF junctions CAR can be enhanced over ET. Therefore, a positive value of $G_{21}$ from dominating CAR as in Fig.\,\ref{crepin}(b) also signals the emergence of equal spin OTE pairing. 
Moreover, recent advances suggest that CAR can also be experimentally obtained from differential conductance measurements \cite{PhysRevLett.93.197003,PhysRevLett.95.027002}. 
Similarly, in SFS junctions, large equal-spin OTE amplitudes have been predicted to emerge at the energies associated with the Andreev bound states, where a finite phase difference $\phi$ and magnetization are  both crucial ingredients \cite{PhysRevB.97.075408}.

\begin{figure}[!t]
  \begin{center}
\includegraphics[width=.95\textwidth]{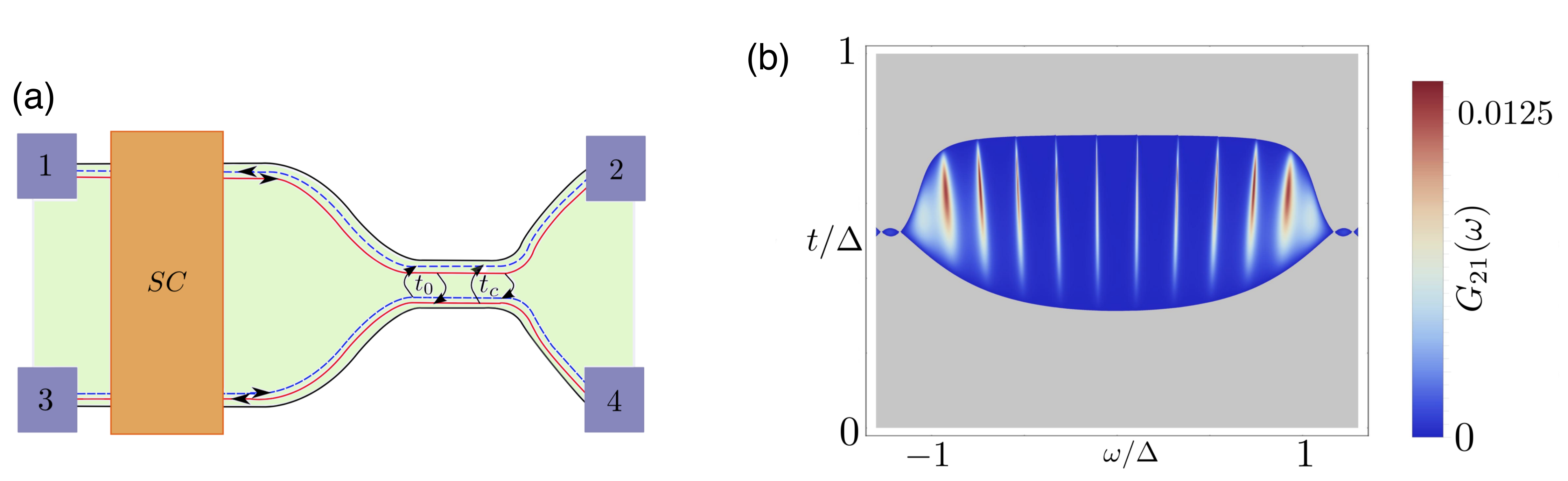} 
  \end{center}
\caption{(a) Two pairs of 1D metallic edges of a 2D TI are coupled through a quantum point contact, with the left region containing proximity-induced conventional superconductivity (SC) and each edge is coupled to separate contacts 1-4. At the quantum point contact, $t_{0}$ and $t_{c}$ correspond to the strength of the two different time-reversal invariant scattering processes across the edges.
(b) Non-local conductance $G_{21}$ between contacts 1-2 as a function of $t_{0}=t_{c}=t$ and frequency $\omega$.
A positive non-local conductance gives dominant CAR, which signals the emergence of equal spin OTE pairing.  
Reprinted figure with permission from [C. Fleckenstein, N. Traverso Ziani, and B. Trauzettel, Phys. Rev. B 97, 134523 (2018)] Copyright (2018) by The American Physical Society. 
}
\label{ziani}
   \end{figure}

It is here important to note that, while we have already discussed methods for generating pair amplitudes in all of the symmetry classes of Table~\ref{tab1}, the only methods we have discussed so far for realizing equal-spin triplet pairing have relied on the presence of a magnetic field. However, it has recently been proposed that equal-spin OTE correlations can also be realized using a quantum point contact connecting the two opposing helical edges of a 2DTI and with proximity-induced conventional spin-singlet $s$-wave superconductivity \cite{PhysRevB.97.134523}, as depicted in Fig.\,\ref{ziani}(a). 
In this setup, the generation of equal-spin OTE pairing is possible due to the presence of spin-conserving backscattering and forward scattering that breaks axial spin symmetry, with both processes being time-reversal invariant across the edges. Interestingly, the authors were able to correlate the CAR process between contacts 1 and 2 in Fig.\,\ref{ziani} with non-local odd-$\omega$ equal-spin triplet pairing .
The authors of Ref.\,\cite{PhysRevB.97.134523} predicted that this effect can be seen in the non-local conductance $G_{21}=(e^{2}/h)(CAR-ET)$, similar to the proposal discussed  previously for the FS and SFS junctions. They also showed that the magnitude of ET per channel can be reduced in this setup, allowing the CAR to dominate over the ET contribution. Thus a positive signal in the non-local conductance, as shown in Fig.\,\ref{ziani}(b), corresponds to the presence of equal-spin OTE. This geometry has recently been fabricated with initial measurements reported \cite{strunz19}, although signatures of OTE still remain to be measured.

We conclude this section by stressing that the superconducting junction geometries discussed here have been fabricated, experimentally, and well-characterized \cite{PhysRevLett.109.186603,Yacoby14,vlad15,Bocquillon17,tkachov19review}. Therefore, all experimental prerequisites for the study of odd-$\omega$ pair correlations in these systems have already been met. 

\section{Odd-$\omega$ pairing and Majorana zero modes}
\label{sec5}
In the previous two sections we discussed the generation of odd-$\omega$ spin-triplet correlations which relied on Rashba SOC in 1D nanowires and helicity in the metallic edges of 2D TIs, respectively. These systems are both highly relevant for the realization of 1D topological superconductivity.
In this section we focus on the deep relationship between odd-$\omega$ pairing and the MZMs which are predicted to emerge as boundary modes in many topological superconductors. In fact, as we will discuss, not only is it possible for the same system to realize both odd-$\omega$ pairing and MZMs, the appearance of MZMs is always accompanied by odd-$\omega$ pairing. This further highlights the fundamental relationship between odd-$\omega$ pairing and topological superconducting phases, and implies that any unambiguous signature of MZMs can also be used to identify odd-$\omega$ pairing.

MZMs, or more generally Majorana fermions, are fermionic particles which are their own antiparticles, i.e.~their creation and annihilation operators are equal, $\gamma=\gamma^{\dagger}$ \cite{majorana}. 
While Majorana fermions were originally discussed in the context of high-energy particle physics, similar states have also been shown to emerge in condensed matter systems. Notably, zero-energy modes with this characteristic, i.e.~MZMs, emerge in the 1D Kitaev model, which consists of a 1D chain of spinless fermions with $p$-wave pairing \cite{kitaev}. More specifically, the Kitaev model exhibits a topological superconducting phase in which two MZMs, $\gamma_{1}$ and $\gamma_{2}$, emerge at the two ends of the system. The magnitudes of the wavefunctions for these modes decay exponentially into the bulk \cite{kitaev}. For finite length chains this implies a finite spatial overlap between the MZMs residing at opposite ends of the system, leading to a  hybridization of the MZMs into non-Majorana quasiparticles of finite energy \cite{PhysRevB.86.085408,PhysRevB.86.180503,PhysRevB.86.220506,PhysRevB.87.024515}.
However, if the two modes $\gamma_{1,2}$ are far enough apart, then the Majorana condition, $\gamma_{i}=\gamma_{i}^{\dagger}$, can be effectively satisfied and the modes also appear at zero energy. 
Since the operators $\gamma_{i}$ for the MZMs are real and highly localized, this implies that, together, a pair of MZMs defines a single highly non-local complex fermion, $c=\left(\gamma_1+i\gamma_2\right)/\sqrt{2}$.

Crucially, the Kitaev model is not just a toy model but holds experimental relevance as an effective model for some superconducting heterostructures \cite{PhysRevLett.105.077001,PhysRevLett.105.177002,PhysRevB.79.161408,PhysRevB.84.195442}. One promising proposal utilizes a combination of 1D nanowires with Rashba SOC, conventional spin-singlet $s$-wave superconductivity, and a magnetic field that drives the system into a topological phase \cite{PhysRevLett.105.077001,PhysRevLett.105.177002}. 
Other proposals for MZMs in 1D combine conventional superconductivity with either metallic edges of 2D TIs \cite{PhysRevB.79.161408}  or 1D chains of magnetic atoms \cite{PhysRevB.84.195442}. 
Recent experiments have measured several predicted signatures of MZMs; however, experimental challenges remain regarding the conclusive discrimination between the topological superconducting phase and the trivial phase and thus the existence of MZMs \cite{Aguadoreview17,LutchynReview08,zhangreview,magnatoms,tkachov19review}. 

To see that odd-$\omega$ pairing and MZMs must be related, we begin by noting that, by inspection of the Majorana condition, $\gamma=\gamma^{\dagger}$ \cite{majorana}, it is clear that the normal ($g$) and anomalous ($f$) propagators for an MZM are the same, $ g(\omega_{m})=f(\omega_{m})$. Furthermore, since they are associated with a quasiparticle at zero energy, we directly get \cite{PhysRevB.92.121404}
\begin{equation}
f(\omega_{m})=\frac{1}{i\omega_{m}}\,
\end{equation}
where $\omega_{m}$ is the Matsubara frequency. Therefore, the pair amplitude $f$ for an isolated MZM is necessarily odd in frequency and divergent at low frequencies $\omega_{m}$, as discussed in Refs.\,\cite{lutchyn16,Takagi18,tamura18}.
With this intrinsic relationship in mind, we now discuss the connection between odd-$\omega$ pairing and MZMs in models describing realistic systems.

\begin{figure}
  \begin{center}
\includegraphics[width=.8\textwidth]{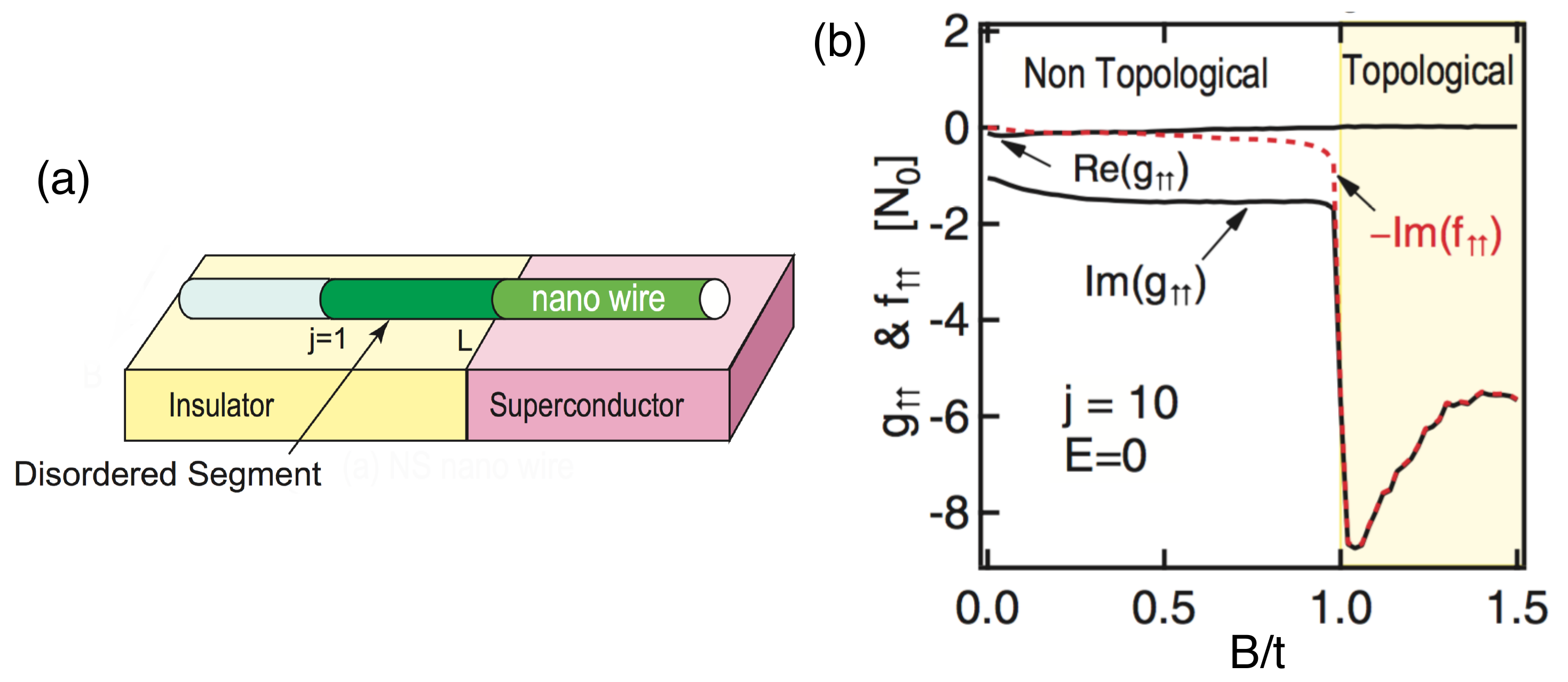} 
  \end{center}
\caption{(a) 1D nanowire with Rasbha SOC and under a Zeeman field $B$ in contact with a superconductor and an insulator. The left region of the nanowire contains a finite disordered section of length $L$. (b) The normal Green's function $g_{\uparrow\uparrow}$ (black solid curve) and  equal-spin pair amplitude $f_{\uparrow\uparrow}$ (red dashed curve) at zero energy ($E=0$) inside of the finite disordered segment (site $j=10$) as a function of the Zeeman field $B$. In the topological phase, marked by the yellow region, $g_{\uparrow\uparrow}$ and $f_{\uparrow\uparrow}$ exhibit a sharp increase due to the MZM. Reprinted figure with permission from [Y. Asano and Y. Tanaka, Phys. Rev. B 87, 104513 (2013)] Copyright (2013) by The American Physical Society. 
}
\label{tanaka1}
   \end{figure}
The relation between MZMs and odd-$\omega$ pairing was initially studied in disordered finite length NS junctions based on nanowires with Rasha SOC in proximity to conventional $s$-wave superconductors, as depicted in Fig.\,\ref{tanaka1}(a) \cite{PhysRevB.87.104513}. The authors showed that the spin-polarized OTE amplitude $f$ and the normal Green's function $g$ in the middle of N exhibit a fast increase as soon as the S region enters the topological phase, as illustrated in Fig.\,\ref{tanaka1}(b). This occurs due to a strong relation between the OTE amplitude and the normal Green's function in the topological phase, which the authors showed to be $f^{r}(x,x';\omega)=i{\rm e}^{i\phi}g^{r}(x,x';\omega)$, where $g^{r}(x,x';\omega)$ is the normal propagator. Furthermore, equal-spin OTE pair correlations were found to survive long distances in N despite the disorder \cite{PhysRevB.87.104513,Takagi18}, as their $s$-wave nature provides a robustness against disorder. This long-distance proximity effect coincides with the appearance of a zero-energy peak in the LDOS, due to the emergence of a MZM at the NS interface \cite{PhysRevB.70.012507,PhysRevB.87.104513,Takagi18}.  

\begin{figure}
  \begin{center}
\includegraphics[width=.8\textwidth]{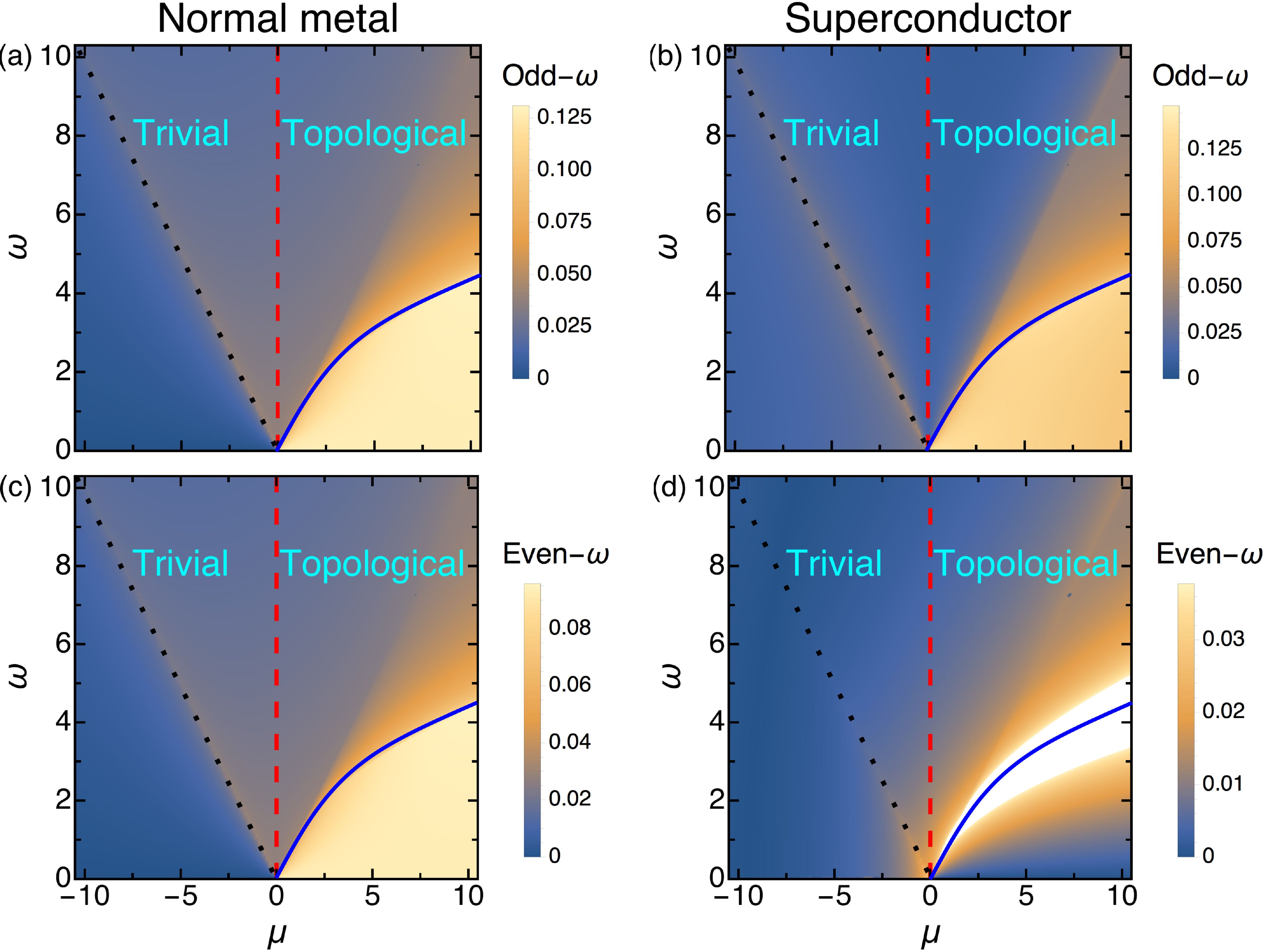} 
  \end{center}
\caption{(a,b) Odd-$\omega$ (OTE) and  even-$\omega$ (ETO) (c,d) pair amplitudes as a function of the chemical potential in S, $\mu$, and frequency for NS Kitaev junctions in both the N and S regions. In general, both pairings coexist but with larger OTE in the topological phase. Vertical red dashed line indicates the topological phase transition, while the solid blue and dotted black  curves show the behavior of the energy gaps in the topological and trivial phases. Adapted from Ref.\,\cite{thanos2019}.}
\label{thanos1}
   \end{figure}
 
As seen above, the odd-$\omega$ pairing in 1D Rashba SOC nanowires is clearly enhanced whenever MZMs are present. However, since MZMs only emerge at interfaces of topological superconducting systems and we already know that OTE pairing can appear at a variety of non-topological interfaces, it is natural to ask whether odd-$\omega$ pairing should be attributed solely due to the emergence of MZMs.  
This issue was recently addressed in ballistic NS and short SNS Kitaev-based junctions following the same scattering Green's function approach used in Sections \ref{sec3}-\ref{sec4} \cite{thanos2019}.  Focusing first on the NS case, the authors of Ref.\,\cite{thanos2019} show that in the N region, ETO and OTE amplitudes (spin-singlet pairing cannot occur in the Kitaev model) are both large in both the trivial and topological phases as they occur simultaneously due to AR, see Fig.\,\ref{thanos1}, although OTE dominates in the topological phase. In the S region the pair amplitudes obtain contributions, both, from the bulk, with ETO symmetry, and the interface, with ETO and OTE symmetries. The OTE amplitude in the S region of NS junctions decays exponentially into the bulk and notably does  not exhibit the $1/|\omega|$ dependence reported in previous works in the presence of an isolated MZM \cite{lutchyn16,Takagi18,tamura18}, which was attributed to the wavefunction of the MZM having a finite width beyond the interface \cite{thanos2019}. 
Moreover, the interface terms arise due to both NR and AR, similarly to Section \ref{sec3}, where AR allows the coexistence of both ETO and OTE pairings, while NR only contributes to OTE pairing. In S, the interface ETO component is always smaller than the OTE pairing due to the features of the ARs, as also seen in Fig.\,\ref{thanos1}. This implies that the MZM induces large OTE correlations \cite{thanos2019}, consistent with the universal spin-triplet superconducting correlations of MZMs found in Ref.\,\cite{PhysRevB.92.014513}.
Despite the fact that these are junctions with $p$-wave superconductors, the coexistence of ETO and OTE amplitudes is mediated by the AR, similar to junctions with $s$-wave superconductors \cite{PhysRevB.92.205424,PhysRevB.96.155426,PhysRevB.98.075425}. 
Since the AR can be obtained from the LDOS or conductance measurements\cite{chang15,doi:10.1063/1.4971394,gulonder,Gazibegovic17,zhang18}, these results can be used to characterize the OTE pairing, provided NRs are reduced e.g.~in high-transparency junctions.

\begin{figure}
  \begin{center}
\includegraphics[width=.75\textwidth]{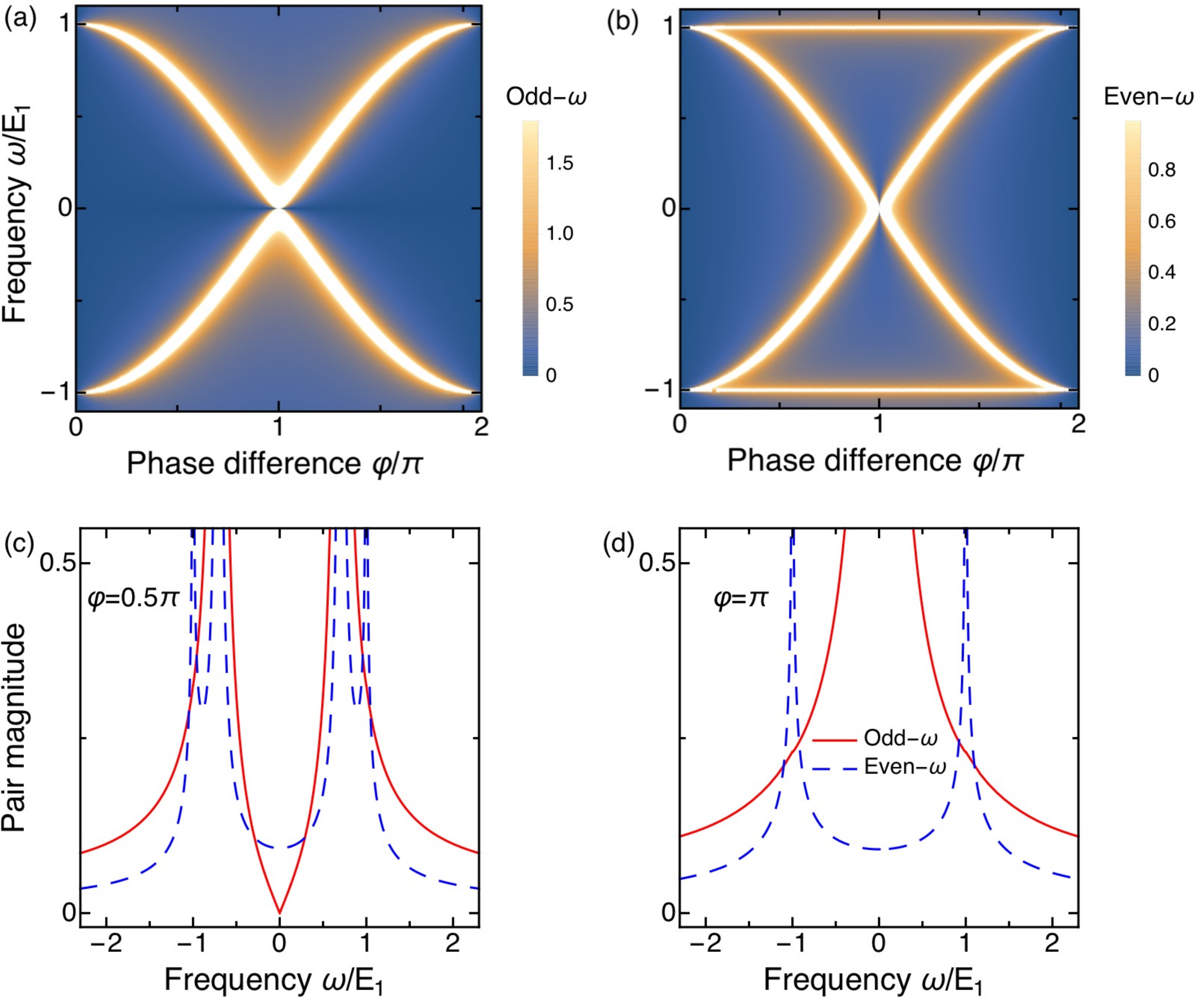} 
  \end{center}
\caption{ (a,b) Even-$\omega$ (ETO) and odd-$\omega$ (OTE) pair magnitudes in the topological phase of short SNS Kitaev junctions as a function of frequency and  phase difference $\phi$. Bright areas in (a,b) coincide with the emergence of topological ABSs. (c,d) Frequency dependence of ETO (dashed blue  curve) and OTE (solid red) magnitudes for fixed $\phi$. Notice that the OTE magnitude in (d) exhibits a sharp peak at zero frequency due to MZMs. Adapted from Ref.\,\cite{thanos2019}.}
\label{thanos2}
   \end{figure}
Next, considering the topological phase of short SNS Kitaev junctions, both ETO and OTE pair amplitudes capture  the emergence of the topological Andreev bound states \cite{thanos2019}, as displayed in Fig.\,\ref{thanos2}.  Exactly at $\phi=\pi$ only the OTE pairing develops a divergence ($\sim1/|\omega|$). This coincides with the zero-frequency crossing at $\phi=\pi$ in topological short SNS junctions which is protected by the conservation of the total fermion parity and signals the emergence of two MZMs at the junction \cite{PhysRevB.53.9371,kitaev,Kwon2004}. Notice how the $\sim1/|\omega|$ behavior of the OTE pairing for SNS junctions is distinct from NS junctions, where the OTE pairing does not diverge at low-frequencies despite the presence of MZMs \cite{thanos2019}. Away from $\phi=\pi$, the OTE amplitude exhibits a linear dependence ($\sim\omega$). This result is consistent with the recent results in Ref.\,\cite{tamura18}, which suggests the OTE amplitude in this system can be written as $f\approx Z/\omega+B\omega$ \cite{tamura18}, where the first (second) term corresponds to the system with (without) MZMs at $\phi=\pi$ (away from $\phi=\pi$).
In fact, the authors of Ref.\,\cite{tamura18} show that $Z$ and $B$ are physical bulk quantities that allow the characterization of the odd-$\omega$ pairs accumulated at the boundary, giving rise to a frequency dependent, or spectral, bulk-boundary correspondence. We conclude by stressing that odd-$\omega$ pairing exists even without MZMs simply due to spatial parity breaking of the order parameter $\Delta(x)$ at interfaces; however, in the presence of MZMs, odd-$\omega$ pairing is significantly enhanced, although not always with a divergent behavior.

There also exist another type of system in which odd-$\omega$ pairing and topology have been shown to meet: double wires with Rashba SOC proximity-coupled to a conventional superconducting substrate \cite{Ebisu16}. This system is particularly interesting because it can enter a topological phase if the induced interwire pairing dominates, corresponding to dominant CARs \cite{PhysRevB.90.045118}. 
In this regime Majorana Kramers pairs, or even parafermions if the system is interacting, can be induced.
The generation of odd-$\omega$ correlations are allowed due to the additional degree of freedom offered by the wire index \cite{Ebisu16,Triola19b}. 
The authors of Ref.\,\cite{Ebisu16} showed that odd-$\omega$ pairing acquires huge values, irrespective of both the spin and spatial symmetries, if topological superconductivity is realized. This device has recently been fabricated and CAR has also been measured \cite{Baba_2018}, which represents an exciting preliminary advance towards the characterization of odd-$\omega$ correlations in this setup.

\subsection{Majorana devices}
\begin{figure}
  \begin{center}
\includegraphics[width=.5\textwidth]{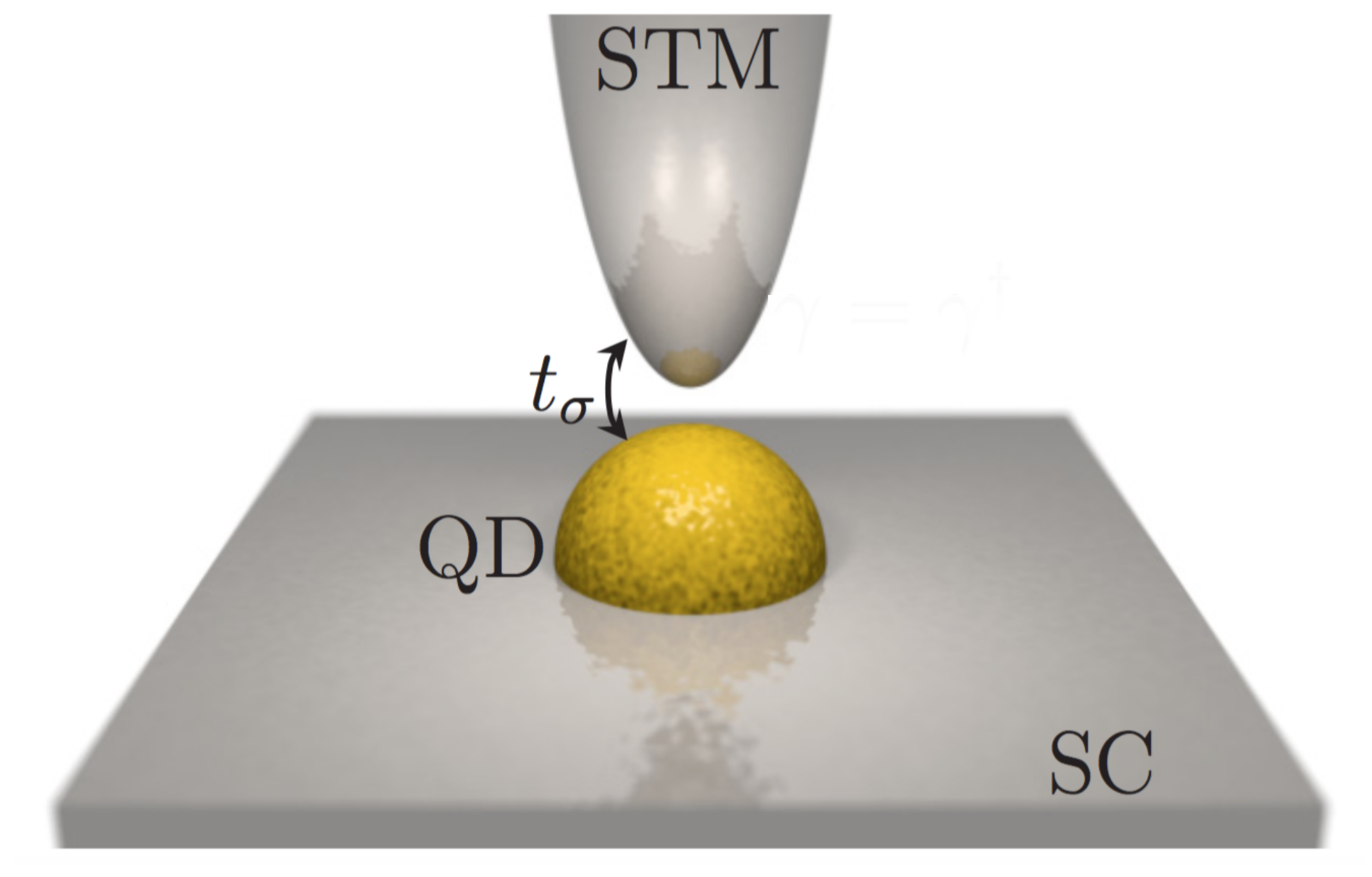} 
  \end{center}
\caption{Majorana STM tip tunnel coupled to a quantum dot (QD) via $t_{\sigma}$ which acquires pair correlations from an unknown superconductor SC. The setup only registers a current flow when the unknown SC has components of equal-spin OTE symmetry. Reprinted figure with permission from [O. Kashuba, B. Sothmann, P. Burset, and B. Trauzettel, Phys. Rev. B 95, 174516 (2017)]
 Copyright (2017) by The American Physical Society.}
\label{kashuba}
   \end{figure}

The close connection between MZM and odd-$\omega$ pairing has recently been utilized to propose devices where odd-$\omega$ pairing is explicitly measured or induced using MZMs.
For example, in Ref.~\cite{PhysRevB.95.174516} it was recently proposed that MZMs can be used to directly probe odd-$\omega$ pairing. Such a protocol, termed Majorana STM and shown schematically in  Fig.\,\ref{kashuba}, takes advantage of the inherently odd-$\omega$ nature of MZMs together with the Josephson effect. In this setup, a single MZM is used as an STM tip. Here the end of a 1D topological superconductor can be used to generate the single MZM, as discussed in the beginning of this section. When placed near the surface of an unknown superconductor, this device would detect a finite Josephson current between the tip and the superconductor if and only if the unknown superconductor possessed equal-spin OTE symmetry, which is the same pair symmetry of MZMs.  
If instead the unknown superconductor has a different pair symmetry, the current flow is completely suppressed due to different symmetries of the involved correlations.

\begin{figure}
  \begin{center}
\includegraphics[width=.9\textwidth]{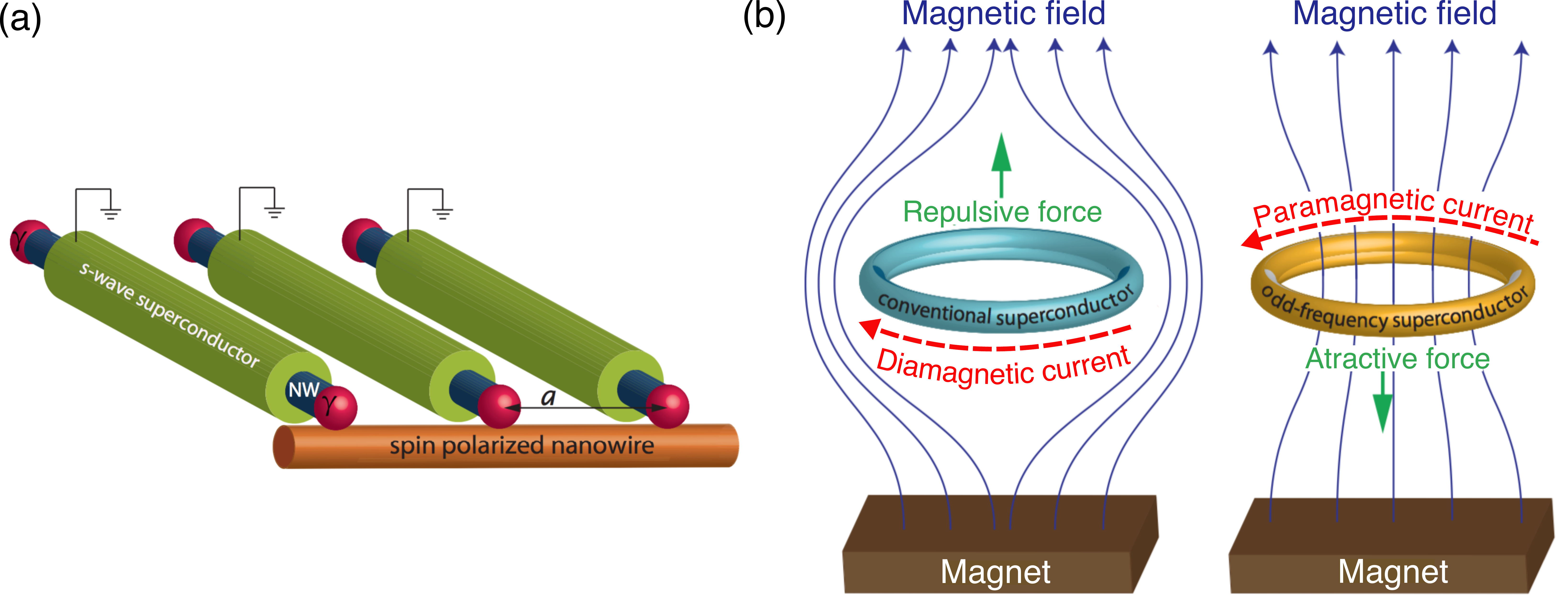} 
  \end{center}
\caption{(a) Array of MZMs (red filled spheres) coupled to a spin polarized nanowire (SPW). The  separation between nearest MZMs is  $a$.
OTE correlations are induced in t he SPW and a 1D odd-$\omega$ bulk superconductor can emerge. (b) Unlike conventional superconductors, the odd-$\omega$ superconductor exhibits a paramagnetic Meissner effect. Reprinted figure with permission from [S. P. Lee, R. M. Lutchyn, and J. Maciejko, Phys. Rev. B 95, 184506 (2017)]
 Copyright (2017) by The American Physical Society.}
\label{lutchyn}
   \end{figure}

Another recently proposed device linking odd-$\omega$ pairing and MZM was recently proposed in Ref.~\cite{lutchyn16}, where a bulk 1D odd-$\omega$ superconductor was predicted to emerge by coupling an array of MZMs to a spin-polarized wire (SPW), as shown schematically in Fig.\,\ref{lutchyn}(a). Since MZMs are spin-polarized, using a SPW simplifies the coupling between MZMs and the wire.
This proposal was motivated by the fact that an isolated MZM exhibits purely odd-$\omega$ pairing, as discussed above, which could be induced by a proximity effect into a different system. For example, if we consider a pair of MZMs with energy splitting $\delta$, with only one MZM coupled to a single quantum dot (QD), the induced pair amplitude in the QD is given by \cite{DushkoMZM}
\begin{equation}
\label{QDMZM}
f(i\omega_{m})=\frac{i}{2}\frac{\omega_{m}\Gamma^{2}}{\omega_{m}^{4}+(\delta^{2}+\epsilon^{2}+|\Gamma|^{2})\omega_{m}^{2}+\delta^{2}\epsilon^{2}}
\end{equation}
where $\epsilon$ is the QD energy level and $\Gamma$ is the coupling strength of a single MZM to the QD. Hence, a MZM induces only OTE pairing, which is a consequence of the QD not having any spatial degrees of freedom. Next, if the single MZM is instead coupled to a SPW, as suggested in Ref.\,\cite{lutchyn16}, the pair amplitude induced in the SPW becomes
\begin{equation}
\label{MZMwire}
f(x,x';i\omega_{m})=-T_{e h} \, \frac{m^{2}}{|p_{0}|^{2}}\,{\rm e}^{i\,{\rm sgn}(\omega_{m})[p_{0}|x|-p_{0}^{*}|x'|]}
\end{equation}
where $p_{0}=\sqrt{2m(\mu+i\omega_{m})/\hbar^{2}}$, $\mu$ the chemical potential in the SPW, $T_{eh}=-(i\omega_{m}/D)\Gamma^{2}$ is odd in frequency, and $D$ is an even function of $\omega_{m}$, whose explicit expression can be found in Ref.\,\cite{lutchyn16}.
Here observe that Eq.\,(\ref{MZMwire}) exhibits translation-invariance breaking due to the MZM, reflected in the exponential term, which mixes electron ($p_{0}$) and hole ($p_{0}^{*}$) wavevectors with different spatial coordinates $x$ and $x'$. This is similar to the situation in Section \ref{sec2} for NS junctions. Still, locally in space, at $x=x'$, the pair amplitude in Eq.\,(\ref{MZMwire}) is purely odd in frequency with OTE symmetry, in agreement with the QD case given by Eq.\,(\ref{QDMZM}). However, non-locally, the parity mixing term results in a single MZM coupled to the SPW inducing a coexistence of ETO and OTE \cite{DushkoMZM}. This can be understood by writing the wavevectors $p_{0}$ in the exponent of Eq.\,(\ref{MZMwire}) for large $\mu$, and then for the exponent we obtain $k_{\rm F}(|x|-|x'|)+i\,\kappa (|x|+|x'|)$ with $\kappa=\omega_{m}/(2\mu)$, $k_{\rm F}=\sqrt{2m\mu/\hbar^{2}}$. This allows the coexistence of ETO and OTE correlations, which decay exponentially with a decay length given by $\kappa^{-1}$ and exhibit oscillations given by $k_{\rm F}$. At low frequencies the decay of both amplitudes is then slow and they spread in space over large distances. Locally in space, however, only OTE correlations survive.

Finally, when a whole array of MZMs is coupled to the SPW, as in Ref.\,\cite{lutchyn16}, a 1D odd-$\omega$ superconductor is realized as each MZM locally induces purely OTE correlations. Non-locally in space, of course, the ETO pairing is still finite but, in general, smaller than OTE pairing \cite{DushkoMZM}. An interesting property of this odd-$\omega$ superconductor is that its Meissner effect has a larger paramagnetic  than diamagnetic contribution. In the paramagnetic Meissner effect the magnetic flux is enhanced instead of screened by the induced supercurrent, as schematically shown in Fig.\,\ref{lutchyn}(b). The authors of Ref.\,\cite{lutchyn16} showed that this paramagnetic contribution implies that the superfluid density is negative for the SPW. 
This view seems to be in agreement with previous studies, where a paramagnetic Meissner effect has been associated with odd-$\omega$ correlations \cite{PhysRevB.64.132507,PhysRevX.5.041021}.

Another example demonstrating the relation between odd-$\omega$ pairing and MZMs was recently investigated in the Sachdev-Ye-Kitaev (SYK) model \cite{PhysRevB.99.024506}. The SYK model consists of $N$ interacting MZMs, where the interactions between MZMs are all-to-all and completely random \cite{PhysRevLett.70.3339,kitaevsyk}. The SYK model can be realized in a system of 1D nanowires as MZMs emerge at their ends \cite{PhysRevB.96.121119}. In Ref.\,\cite{PhysRevB.99.024506} a non-interacting single-state QD was considered randomly coupled to the MZMs and solved in the large $N$ limit, for low frequencies. Interestingly, the  Green's function in the QD acquires anomalous terms given by $f(i\omega_{m})=[\lambda^{2}G_{\gamma}(i\omega_{m})]/[\epsilon^{2}-i\omega_{m}[i\omega_{m}-2\lambda^{2}G_{\gamma}(i\omega_{m})]]$, where $G_{\gamma}=-i\pi^{1/4}{\rm sgn}(\omega_{m}(J|\omega_{m}|)^{-1/2})$, $\lambda$  characterizes the  variance of the couplings between the MZMs in the SYK model and the QD,  $J$ describes the variance of the interactions between MZMs, and $\epsilon$ is the QD energy. 
Thus the anomalous propagator $f$ hosts an odd-$\omega$ pair amplitude. This behavior is similar to the pair amplitude for a MZM when coupled to a QD, as we can see by comparing it with Eq.\,(\ref{QDMZM}). 
Moreover, the author of Ref.\,\cite{PhysRevB.99.024506} stresses that the absence of a gap in the presence of this pairing indicates the emergence of gapless superconductivity in zero dimensions \cite{gaplesssc,PhysRevLett.9.315,PhysRevLett.10.96}.

To summarize this section, we have seen that, although odd-$\omega$ pairing can be present in the topologically trivial phase, in which MZMs are absent, when MZMs are present the system necessarily hosts odd-$\omega$ correlations and the magnitudes of these correlations can be greatly enhanced by contributions related to the MZMs. Therefore, any signature of MZMs directly implies the emergence of odd-$\omega$ pairing \cite{Aguadoreview17,LutchynReview08,zhangreview,magnatoms,tkachov19review}.

\section{Concluding remarks}
\label{sec7}
In this work we have reviewed some recent advances in the study of odd-$\omega$ pair correlations in 1D systems with proximity-induced superconductivity. In particular, we focused on systems that hold relevance for 1D topological superconductivity, such as nanowires with Rashba SOC and the metallic edges of 2D TIs. We also highlighted the deep connection between odd-$\omega$ pairing in these systems and the Majorana zero modes (MZMs) which emerge at their boundaries the topological phase.  
 
First, we provided a basic introduction to the symmetry classification of Cooper pairs based on the constraints imposed by Fermi-Dirac statistics. For the systems considered in this review, four symmetry classes are possible: even-$\omega$ spin-singlet even-parity (ESE); even-$\omega$ spin-triplet odd-parity (ETO); odd-$\omega$ spin-singlet odd-parity (OSO); and odd-$\omega$ spin-triplet even-parity (OTE). We then proceeded to illustrate the way in which the odd-$\omega$ amplitudes can emerge in spin-degenerate 1D systems forming NS and SNS junctions. Importantly, the presence of an interface leads to frequency-dependent pair amplitudes, through Andreev reflection processes, with the breaking of spatial translation-invariance facilitating a mixing of the spatial parities within these amplitudes. This combines to generate odd-$\omega$ pairing, even in the absence of spin-dependent fields such as magnetism. 

Next, we showed that for nanowires with strong Rashba SOC and proximity-coupled to conventional superconductors, the presence of the SOC term leads to odd-$\omega$ mixed-spin triplet correlations in the nanowire, in addition to the odd-$\omega$ spin-singlet amplitudes induced already in the spin-degenerate case. In fact, with strong SOC, pair amplitudes belonging to all four of the allowed symmetry classes for Cooper pairs can be realized in these systems. The relationship between these induced pair amplitudes and the normal and Andreev reflection coefficients was highlighted, exposing that the proximity-induced pair amplitudes can be characterized by conductance measurements. We then discussed the case of odd-$\omega$ pairing in the helical edges of 2D TIs coupled to conventional superconductors, pointing out the similarities and differences between this case and that of Rashba SOC nanowires. Similar to the situation in Rashba SOC nanowires, all of the possible symmetry classes can be realized in these helical edges, and their amplitudes can be characterized, directly, using conductance measurements. 
Notably, for helical edge modes, by engineering a quantum point contact between the two edges, also odd-$\omega$ equal-spin triplet pairing can be generated without the use of magnetic materials. This will undoubtedly motivate future work towards applications in superconducting spintronics without the necessity to engineer complicated magnetic structures. 

We also elaborated on the relationship between odd-$\omega$ pairing and MZMs. After a brief introduction to MZMs, we noted that the anomalous propagator, and thus the pair amplitude, for a MZM is simply $f = g = 1/\omega$, since the Majorana creation and annihilation operators are equal to one another and the MZM is pinned to zero energy. Thus pair correlations associated with MZMs are necessarily odd-$\omega$ amplitudes. We then discussed various works further investigating this connection in models of real systems, including proposals to utilize MZM to detect or engineer odd-$\omega$ superconductors.
 
Throughout this review we have also discussed various experimental methods for characterizing odd-$\omega$ pair amplitudes. Notably, these included: zero-bias peaks in the local density of states, direct measurements of the Andreev reflection coefficients using the conductance, as well as Majorana scanning tunneling microscopy and the paramagnetic Meissner effect. While each of these signatures provides a clear cut way of discerning whether or not odd-$\omega$ pairing exists in the system, they are all system-dependent. As far as we know, none of these tools provides an unambiguous direct measurement of all kinds of odd-$\omega$ pair amplitudes in a generic system. Therefore, further studies on direct probes of the symmetries of Cooper pairs are still highly desirable, where dynamical protocols may be necessary to resolve the frequency-dependence. Moreover, as stressed at the end of Sec.~\ref{sec4}, the superconducting junction geometries discussed there are already under active investigation and all experimental prerequisites for the study of odd-$\omega$ pair correlations in those systems have been met. Given that these systems, as well as most junctions discussed throughout this review, possess a variety of interesting properties that can be easily tuned, we expect that the search for applications of odd-$\omega$ pairing in these 1D systems will continue to be an active area of investigation for the foreseeable future.

\section*{Acknowledgments}
We thank R. Aguado, O. Awoga, A. Balatsky, M. Benito, A. Bouhon,  G. Burkard, P. Burset, P. Dutta,  Y. Fukaya, L. Komendova, D. Kuzmanovski, 	T. L\"{o}thman, B. Lu,  M. Mashkoori, S. Nakosai, F. Parhizgar, C. Reeg, J. Schmidt,  B. Sothmann, S. Tamura, Y. Tanaka, G. Tkachov,  A. Tsintzis, and S. Suzuki  for interesting discussions. This work was made possible by support from the Swedish Research Council (Vetenskapsr\aa det, 2018-03488), the G\"{o}ran Gustafsson Foundation, the Knut and Alice Wallenberg Foundation through the Wallenberg Academy Fellows program, and the European Research Council (ERC) under the European Union's Horizon 2020 research and innovation programme (ERC-2017-StG-757553).

\section*{Author contribution statement}
J.C. conceived the idea and wrote the manuscript with input from C. T. and A.B.S. All authors discussed the results, provided scientific insight, and contributed to the final version of the manuscript. 

\bibliographystyle{epj}

\bibliography{biblio}

\begin{thebibliography}{236}

\bibitem{kamerlingh1911resistance}
H.~Kamerlingh~Onnes, Commun. Phys. Lab. Univ. Leiden, b \textbf{120} (1911)

\bibitem{PhysRev.108.1175}
J.~Bardeen, L.N. Cooper, J.R. Schrieffer, Phys. Rev. \textbf{108}, 1175 (1957)

\bibitem{RevModPhys.72.969}
C.C. Tsuei, J.R. Kirtley, Rev. Mod. Phys. \textbf{72}, 969 (2000)

\bibitem{RevModPhys.75.657}
A.P. Mackenzie, Y.~Maeno, Rev. Mod. Phys. \textbf{75}, 657 (2003)

\bibitem{eliashberg1960interactions}
G.~Eliashberg, Sov. Phys. JETP \textbf{11}, 696 (1960)

\bibitem{abrikosov2012methods}
A.A. Abrikosov, L.P. Gorkov, I.E. Dzyaloshinski, \emph{Methods of quantum field
  theory in statistical physics} (Courier Corporation, 2012)

\bibitem{mahan2013many}
G.D. Mahan, \emph{Many-particle physics} (Springer Science \& Business Media,
  2013)

\bibitem{bere74}
V.L. Berezinskii, JETP Lett. \textbf{20}, 287 (1974)

\bibitem{PhysRevLett.66.1533}
T.R. Kirkpatrick, D.~Belitz, Phys. Rev. Lett. \textbf{66}, 1533 (1991)

\bibitem{PhysRevB.46.8393}
D.~Belitz, T.R. Kirkpatrick, Phys. Rev. B \textbf{46}, 8393 (1992)

\bibitem{PhysRevB.45.13125}
A.~Balatsky, E.~Abrahams, Phys. Rev. B \textbf{45}, 13125 (1992)

\bibitem{PhysRevB.47.513}
E.~Abrahams, A.~Balatsky, J.R. Schrieffer, P.B. Allen, Phys. Rev. B
  \textbf{47}, 513 (1993)

\bibitem{PhysRevB.46.10812}
V.J. Emery, S.~Kivelson, Phys. Rev. B \textbf{46}, 10812 (1992)

\bibitem{PhysRevB.48.7445}
A.V. Balatsky, J.~Bon\v{c}a, Phys. Rev. B \textbf{48}, 7445 (1993)

\bibitem{PhysRevB.47.6157}
N.~Bulut, D.J. Scalapino, S.R. White, Phys. Rev. B \textbf{47}, 6157 (1993)

\bibitem{PhysRevLett.70.2960}
P.~Coleman, E.~Miranda, A.~Tsvelik, Phys. Rev. Lett. \textbf{70}, 2960 (1993)

\bibitem{BALATSKY1994363}
A.~Balatsky, E.~Abrahams, D.~Scalapino, J.~Schrieffer, Physica B: Condens.
  Matter \textbf{199-200}, 363  (1994)

\bibitem{PhysRevB.49.8955}
P.~Coleman, E.~Miranda, A.~Tsvelik, Phys. Rev. B \textbf{49}, 8955 (1994)

\bibitem{PhysRevB.52.1271}
E.~Abrahams, A.~Balatsky, D.J. Scalapino, J.R. Schrieffer, Phys. Rev. B
  \textbf{52}, 1271 (1995)

\bibitem{PhysRevB.52.15649}
E.~Abrahams, A.~Balatsky, J.R. Schrieffer, P.B. Allen, Phys. Rev. B
  \textbf{52}, 15649 (1995)

\bibitem{PhysRevLett.74.1004}
A.V. Balatsky, E.~Abrahams, Phys. Rev. Lett. \textbf{74}, 1004 (1995)

\bibitem{PhysRevLett.74.1653}
P.~Coleman, E.~Miranda, A.~Tsvelik, Phys. Rev. Lett. \textbf{74}, 1653 (1995)

\bibitem{0953-8984-9-2-002}
P.~Coleman, A.~Georges, A.M. Tsvelik, J. Phys.: Condens. Matter \textbf{9}, 345
  (1997)

\bibitem{PhysRevB.59.R713}
M.~Vojta, E.~Dagotto, Phys. Rev. B \textbf{59}, R713 (1999)

\bibitem{PhysRevB.60.3485}
D.~Belitz, T.R. Kirkpatrick, Phys. Rev. B \textbf{60}, 3485 (1999)

\bibitem{doi:10.1143/JPSJ.69.2229}
K.~Hashimoto, J. Phys. Soc. Jpn. \textbf{69}, 2229 (2000)

\bibitem{PhysRevB.64.132507}
K.~Hashimoto, Phys. Rev. B \textbf{64}, 132507 (2001)

\bibitem{doi:10.1143/JPSJ.72.2914}
Y.~Fuseya, H.~Kohno, K.~Miyake, J. Phys. Soc. Jpn. \textbf{72}, 2914 (2003)

\bibitem{PhysRevB.77.144513}
H.~Aizawa, K.~Kuroki, Y.~Tanaka, Phys. Rev. B \textbf{77}, 144513 (2008)

\bibitem{PhysRevB.79.132502}
D.~Solenov, I.~Martin, D.~Mozyrsky, Phys. Rev. B \textbf{79}, 132502 (2009)

\bibitem{PhysRevB.79.174507}
K.~Shigeta, S.~Onari, K.~Yada, Y.~Tanaka, Phys. Rev. B \textbf{79}, 174507
  (2009)

\bibitem{doi:10.1143/JPSJ.78.123710}
T.~Hotta, J. Phys. Soc. Jpn. \textbf{78}, 123710 (2009)

\bibitem{PhysRevB.83.140509}
K.~Shigeta, Y.~Tanaka, K.~Kuroki, S.~Onari, H.~Aizawa, Phys. Rev. B
  \textbf{83}, 140509 (2011)

\bibitem{doi:10.1143/JPSJ.80.054702}
H.~Kusunose, Y.~Fuseya, K.~Miyake, J. Phys. Soc. Jpn. \textbf{80}, 054702
  (2011)

\bibitem{doi:10.1143/JPSJ.80.044711}
H.~Kusunose, Y.~Fuseya, K.~Miyake, J. Phys. Soc. Jpn. \textbf{80}, 044711
  (2011)

\bibitem{doi:10.1143/JPSJ.81.033702}
M.~Matsumoto, M.~Koga, H.~Kusunose, J. Phys. Soc. Jpn. \textbf{81}, 033702
  (2012)

\bibitem{doi:10.1143/JPSJ.81.123701}
Y.~Yanagi, Y.~Yamashita, K.~Ueda, J. Phys. Soc. Jpn. \textbf{81}, 123701 (2012)

\bibitem{PhysRevB.85.174528}
H.~Kusunose, M.~Matsumoto, M.~Koga, Phys. Rev. B \textbf{85}, 174528 (2012)

\bibitem{PhysRevB.85.224509}
K.~Shigeta, S.~Onari, Y.~Tanaka, Phys. Rev. B \textbf{85}, 224509 (2012)

\bibitem{doi:10.1143/JPSJS.81SB.SB015}
T.~Harada, Y.~Fuseya, K.~Miyake, J. Phys. Soc. Jpn. \textbf{81}, SB015 (2012)

\bibitem{PhysRevLett.110.107005}
F.L. Pratt, T.~Lancaster, S.J. Blundell, C.~Baines, Phys. Rev. Lett.
  \textbf{110}, 107005 (2013)

\bibitem{doi:10.7566/JPSJ.82.104702}
K.~Shigeta, S.~Onari, Y.~Tanaka, J. Phys. Soc. Jpn. \textbf{82}, 104702 (2013)

\bibitem{PhysRevLett.112.167204}
S.~Hoshino, Y.~Kuramoto, Phys. Rev. Lett. \textbf{112}, 167204 (2014)

\bibitem{PhysRevB.90.115154}
S.~Hoshino, Phys. Rev. B \textbf{90}, 115154 (2014)

\bibitem{doi:10.7566/JPSJ.83.123704}
H.~Funaki, H.~Shimahara, J. Phys. Soc. Jpn. \textbf{83}, 123704 (2014)

\bibitem{PhysRevLett.115.036404}
J.~Otsuki, Phys. Rev. Lett. \textbf{115}, 036404 (2015)

\bibitem{KASHIWAGI201629}
M.~Kashiwagi, M.~Kato, Phys. Procedia \textbf{81}, 29  (2016), proceedings of
  the 28th International Symposium on Superconductivity (ISS 2015) Nov. 16-18,
  2015, Tokyo, Japan

\bibitem{PhysRevB.93.224511}
S.~Hoshino, K.~Yada, Y.~Tanaka, Phys. Rev. B \textbf{93}, 224511 (2016)

\bibitem{PhysRevB.71.094513}
Y.~Tanaka, S.~Kashiwaya, T.~Yokoyama, Phys. Rev. B \textbf{71}, 094513 (2005)

\bibitem{PhysRevB.72.140503}
Y.~Tanaka, Y.~Asano, A.A. Golubov, S.~Kashiwaya, Phys. Rev. B \textbf{72},
  140503 (2005)

\bibitem{PhysRevLett.98.037003}
Y.~Tanaka, A.A. Golubov, Phys. Rev. Lett. \textbf{98}, 037003 (2007)

\bibitem{PhysRevLett.99.037005}
Y.~Tanaka, A.A. Golubov, S.~Kashiwaya, M.~Ueda, Phys. Rev. Lett. \textbf{99},
  037005 (2007)

\bibitem{Eschrig2007}
M.~Eschrig, T.~L{\"o}fwander, T.~Champel, J.C. Cuevas, J.~Kopu, G.~Sch{\"o}n,
  J. Low Temp. Phys. \textbf{147}, 457 (2007)

\bibitem{PhysRevB.76.054522}
Y.~Tanaka, Y.~Tanuma, A.A. Golubov, Phys. Rev. B \textbf{76}, 054522 (2007)

\bibitem{PhysRevLett.86.4096}
F.S. Bergeret, A.F. Volkov, K.B. Efetov, Phys. Rev. Lett. \textbf{86}, 4096
  (2001)

\bibitem{Kadigrobov01}
{Kadigrobov, A.}, {Shekhter, R. I.}, {Jonson, M.}, Europhys. Lett. \textbf{54},
  394 (2001)

\bibitem{longrangeExp}
V.T. Petrashov, V.N. Antonov, S.V. Maksimov, R.S. Sha\v{l}kha\v{l}darov, Pis'ma
  Zh. Eksp. Teor. Fiz. \textbf{59}, 523 (1994)

\bibitem{PhysRevB.58.R11872}
M.~Giroud, H.~Courtois, K.~Hasselbach, D.~Mailly, B.~Pannetier, Phys. Rev. B
  \textbf{58}, R11872 (1998)

\bibitem{Keizer06}
R.S. Keizer, S.T.B. Goennenwein, T.M. Klapwijk, G.~Miao, G.~Xiao, A.~Gupta,
  Nature \textbf{439}, 825 (2006)

\bibitem{PhysRevLett.104.137002}
T.S. Khaire, M.A. Khasawneh, W.P. Pratt, N.O. Birge, Phys. Rev. Lett.
  \textbf{104}, 137002 (2010)

\bibitem{wang10}
J.~Wang, M.~Singh, M.~Tian, N.~Kumar, B.~Liu, C.~Shi, J.K. Jain, N.~Samarth,
  T.E. Mallouk, M.H.W. Chan, Nat. Phys. \textbf{6}, 389 (2010)

\bibitem{Robinson59}
J.W.A. Robinson, J.D.S. Witt, M.G. Blamire, Science \textbf{329}, 59 (2010)

\bibitem{PhysRevB.82.100501}
M.S. Anwar, F.~Czeschka, M.~Hesselberth, M.~Porcu, J.~Aarts, Phys. Rev. B
  \textbf{82}, 100501 (2010)

\bibitem{Cirillo_2017}
C.~Cirillo, S.~Voltan, E.A. Ilyina, J.M. Hern{\'{a}}ndez,
  A.~Garc{\'{\i}}a-Santiago, J.~Aarts, C.~Attanasio, New J. Phys. \textbf{19},
  023037 (2017)

\bibitem{PROSHIN2018359}
Y.~Proshin, M.~Avdeev, J. Magn. Magn. Mater. \textbf{459}, 359  (2018), the
  selected papers of Seventh Moscow International Symposium on Magnetism
  (MISM-2017)

\bibitem{PhysRevB.97.100502}
M.V. Avdeev, Y.N. Proshin, Phys. Rev. B \textbf{97}, 100502 (2018)

\bibitem{BELZIG19991251}
W.~Belzig, F.K. Wilhelm, C.~Bruder, G.~Schon, A.D. Zaikin, Superlattices
  Microstruct. \textbf{25}, 1251  (1999)

\bibitem{PhysRevB.62.11377}
A.~Buzdin, Phys. Rev. B \textbf{62}, 11377 (2000)

\bibitem{PhysRevB.62.11846}
M.L. Kuli\ifmmode~\acute{c}\else \'{c}\fi{}, M.~Endres, Phys. Rev. B
  \textbf{62}, 11846 (2000)

\bibitem{PhysRevLett.86.308}
M.~Zareyan, W.~Belzig, Y.V. Nazarov, Phys. Rev. Lett. \textbf{86}, 308 (2001)

\bibitem{PhysRevB.68.064513}
F.S. Bergeret, A.F. Volkov, K.B. Efetov, Phys. Rev. B \textbf{68}, 064513
  (2003)

\bibitem{PhysRevLett.88.047003}
D.~Huertas-Hernando, Y.V. Nazarov, W.~Belzig, Phys. Rev. Lett. \textbf{88},
  047003 (2002)

\bibitem{PhysRevLett.90.137003}
M.~Eschrig, J.~Kopu, J.C. Cuevas, G.~Sch\"on, Phys. Rev. Lett. \textbf{90},
  137003 (2003)

\bibitem{Fominov2003}
Y.V. Fominov, A.A. Golubov, M.Y. Kupriyanov, JETP Lett. \textbf{77}, 510 (2003)

\bibitem{RevModPhys.77.935}
A.I. Buzdin, Rev. Mod. Phys. \textbf{77}, 935 (2005)

\bibitem{RevModPhys.77.1321}
F.S. Bergeret, A.F. Volkov, K.B. Efetov, Rev. Mod. Phys. \textbf{77}, 1321
  (2005)

\bibitem{PhysRevLett.98.077003}
V.~Braude, Y.V. Nazarov, Phys. Rev. Lett. \textbf{98}, 077003 (2007)

\bibitem{PhysRevLett.98.107002}
Y.~Asano, Y.~Tanaka, A.A. Golubov, Phys. Rev. Lett. \textbf{98}, 107002 (2007)

\bibitem{PhysRevB.75.134510}
T.~Yokoyama, Y.~Tanaka, A.A. Golubov, Phys. Rev. B \textbf{75}, 134510 (2007)

\bibitem{PhysRevB.98.161408}
S.Y. Hwang, P.~Burset, B.~Sothmann, Phys. Rev. B \textbf{98}, 161408 (2018)

\bibitem{PhysRevB.64.134506}
F.S. Bergeret, A.F. Volkov, K.B. Efetov, Phys. Rev. B \textbf{64}, 134506
  (2001)

\bibitem{PhysRevX.5.041021}
A.~Di~Bernardo, Z.~Salman, X.L. Wang, M.~Amado, M.~Egilmez, M.G. Flokstra,
  A.~Suter, S.L. Lee, J.H. Zhao, T.~Prokscha et~al., Phys. Rev. X \textbf{5},
  041021 (2015)

\bibitem{PhysRevB.92.014508}
M.~Alidoust, K.~Halterman, O.T. Valls, Phys. Rev. B \textbf{92}, 014508 (2015)

\bibitem{bernardo15}
A.D. Bernardo, S.~Diesch, Y.~Gu, J.~Linder, G.~Divitini, C.~Ducati, E.~Scheer,
  M.~Blamire, J.~Robinson, Nat. Commun. \textbf{6}, 8053 (2015)

\bibitem{RevModPhys.63.239}
M.~Sigrist, K.~Ueda, Rev. Mod. Phys. \textbf{63}, 239 (1991)

\bibitem{Maeno94}
A.P. Higginbotham, S.M. Albrecht, G.~Kirsanskas, W.~Chang, F.~Kuemmeth,
  P.~Krogstrup, T.S.J.J. Nyg{\aa}rd, K.~Flensberg, C.M. Marcus, Nature
  \textbf{372}, 532 (1994)

\bibitem{PhysRevLett.80.3129}
H.~Tou, Y.~Kitaoka, K.~Ishida, K.~Asayama, N.~Kimura, Y.~\={O}nuki,
  E.~Yamamoto, Y.~Haga, K.~Maezawa, Phys. Rev. Lett. \textbf{80}, 3129 (1998)

\bibitem{PhysRevLett.107.077003}
S.~Kashiwaya, H.~Kashiwaya, H.~Kambara, T.~Furuta, H.~Yaguchi, Y.~Tanaka,
  Y.~Maeno, Phys. Rev. Lett. \textbf{107}, 077003 (2011)

\bibitem{LinderNat15}
J.~Linder, J.W.A. Robinson, Nat. Phys. \textbf{11}, 307 (2015)

\bibitem{7870d3ff91ed485fa3e55e901ff81c80}
M.~Eschrig, Phys. Today \textbf{64}, 43 (2011)

\bibitem{EschrigNat15}
M.~Eschrig, T.~L\"{o}fwander, Nat. Phys. \textbf{4}, 138 (2008)

\bibitem{0034-4885-78-10-104501}
M.~Eschrig, Rep. Prog. Phys. \textbf{78}, 104501 (2015)

\bibitem{PhysRevLett.100.096407}
L.~Fu, C.L. Kane, Phys. Rev. Lett. \textbf{100}, 096407 (2008)

\bibitem{PhysRevB.79.161408}
L.~Fu, C.L. Kane, Phys. Rev. B \textbf{79}, 161408 (2009)

\bibitem{PhysRevLett.105.077001}
R.M. Lutchyn, J.D. Sau, S.~Das~Sarma, Phys. Rev. Lett. \textbf{105}, 077001
  (2010)

\bibitem{PhysRevLett.105.177002}
Y.~Oreg, G.~Refael, F.~von Oppen, Phys. Rev. Lett. \textbf{105}, 177002 (2010)

\bibitem{kitaev}
A.Y. Kitaev, Phys. Usp. \textbf{44}, 131 (2001)

\bibitem{RevModPhys.80.1083}
C.~Nayak, S.H. Simon, A.~Stern, M.~Freedman, S.~Das~Sarma, Rev. Mod. Phys.
  \textbf{80}, 1083 (2008)

\bibitem{Sarma:16}
S.D. Sarma, M.~Freedman, C.~Nayak, Npj Quantum Information \textbf{1}, 15001
  (2015)

\bibitem{Aguadoreview17}
R.~Aguado, La Rivista del Nuovo Cimento \textbf{40}, 523 (2017)

\bibitem{LutchynReview08}
R.M. Lutchyn, E.P.A.M. Bakkers, L.P. Kouwenhoven, P.~Krogstrup, C.M. Marcus,
  Y.~Oreg, Nat. Rev. Mater. \textbf{3}, 52 (2018)

\bibitem{zhangreview}
H.~Zhang, D.E. Liu, M.~Wimmer, L.P. Kouwenhoven, arXiv:1905.07882  (2019)

\bibitem{tkachov19review}
D.~Culcer, A.C. Keser, Y.~Li, G.~Tkachov, arXiv:1907.10058  (2019)

\bibitem{PhysRevB.86.075410}
T.~Yokoyama, Phys. Rev. B \textbf{86}, 075410 (2012)

\bibitem{PhysRevB.86.144506}
A.M. Black-Schaffer, A.V. Balatsky, Phys. Rev. B \textbf{86}, 144506 (2012)

\bibitem{PhysRevB.87.220506}
A.M. Black-Schaffer, A.V. Balatsky, Phys. Rev. B \textbf{87}, 220506 (2013)

\bibitem{PhysRevB.92.205424}
P.~Burset, B.~Lu, G.~Tkachov, Y.~Tanaka, E.M. Hankiewicz, B.~Trauzettel, Phys.
  Rev. B \textbf{92}, 205424 (2015)

\bibitem{Lu_2015}
B.~Lu, P.~Burset, K.~Yada, Y.~Tanaka, Supercond. Sci. Technol. \textbf{28},
  105001 (2015)

\bibitem{PhysRevB.92.100507}
F.~Cr\'epin, P.~Burset, B.~Trauzettel, Phys. Rev. B \textbf{92}, 100507 (2015)

\bibitem{PhysRevB.96.155426}
J.~Cayao, A.M. Black-Schaffer, Phys. Rev. B \textbf{96}, 155426 (2017)

\bibitem{PhysRevB.96.174509}
D.~Kuzmanovski, A.M. Black-Schaffer, Phys. Rev. B \textbf{96}, 174509 (2017)

\bibitem{bo2016}
B.~Lu, Y.~Tanaka, Phil. Trans. R. Soc. A \textbf{376}, 20150246 (2018)

\bibitem{PhysRevB.97.075408}
F.~Keidel, P.~Burset, B.~Trauzettel, Phys. Rev. B \textbf{97}, 075408 (2018)

\bibitem{PhysRevLett.120.037701}
D.~Breunig, P.~Burset, B.~Trauzettel, Phys. Rev. Lett. \textbf{120}, 037701
  (2018)

\bibitem{PhysRevB.97.134523}
C.~Fleckenstein, N.T. Ziani, B.~Trauzettel, Phys. Rev. B \textbf{97}, 134523
  (2018)

\bibitem{PhysRevB.92.134512}
C.R. Reeg, D.L. Maslov, Phys. Rev. B \textbf{92}, 134512 (2015)

\bibitem{Ebisu16}
H.~Ebisu, B.~Lu, J.~Klinovaja, Y.~Tanaka, Prog. Theor. Exp. Phys.
  \textbf{2016}, 083I01 (2016)

\bibitem{PhysRevB.95.184518}
I.V. Bobkova, A.M. Bobkov, Phys. Rev. B \textbf{95}, 184518 (2017)

\bibitem{PhysRevB.98.075425}
J.~Cayao, A.M. Black-Schaffer, Phys. Rev. B \textbf{98}, 075425 (2018)

\bibitem{PhysRevB.99.184501}
S.~Tamura, Y.~Tanaka, Phys. Rev. B \textbf{99}, 184501 (2019)

\bibitem{PhysRevB.90.220501}
B.~Sothmann, S.~Weiss, M.~Governale, J.~K\"onig, Phys. Rev. B \textbf{90},
  220501 (2014)

\bibitem{PhysRevB.93.201402}
P.~Burset, B.~Lu, H.~Ebisu, Y.~Asano, Y.~Tanaka, Phys. Rev. B \textbf{93},
  201402 (2016)

\bibitem{PhysRevB.94.094518}
C.~Triola, A.V. Balatsky, Phys. Rev. B \textbf{94}, 094518 (2016)

\bibitem{triola17}
C.~Triola, A.V. Balatsky, Phys. Rev. B \textbf{95}, 224518 (2017)

\bibitem{PhysRevB.88.104514}
A.M. Black-Schaffer, A.V. Balatsky, Phys. Rev. B \textbf{88}, 104514 (2013)

\bibitem{PhysRevB.92.094517}
L.~Komendov\'a, A.V. Balatsky, A.M. Black-Schaffer, Phys. Rev. B \textbf{92},
  094517 (2015)

\bibitem{PhysRevB.92.224508}
Y.~Asano, A.~Sasaki, Phys. Rev. B \textbf{92}, 224508 (2015)

\bibitem{Samokhvalov_2016}
A.V. Samokhvalov, A.S. Melnikov, A.I. Buzdin, Physics-Uspekhi \textbf{59}, 571
  (2016)

\bibitem{Golubov2011x}
A.A. Golubov, Y.~Tanaka, Y.~Asano, Y.~Tanuma, \emph{Odd-Frequency Pairing in
  Superconducting Heterostructures} (Springer Berlin Heidelberg, Berlin,
  Heidelberg, 2011)

\bibitem{Tanaka2018}
Y.~Tanaka, S.~Tamura, Journal of Low Temperature Physics \textbf{191}, 61
  (2018)

\bibitem{Nagaosa12}
Y.~Tanaka, M.~Sato, N.~Nagaosa, J. Phys. Soc. Jpn. \textbf{81}, 011013 (2012)

\bibitem{Balatsky2017}
J.~Linder, A.V. Balatsky, arXiv:1709.03986  (2017)

\bibitem{zagoskin}
A.~Zagoskin, \emph{Quantum Theory of Many-Body Systems: Techniques and
  Applications} (Springer, 2014)

\bibitem{PhysRev.175.559}
W.L. McMillan, Phys. Rev. \textbf{175}, 559 (1968)

\bibitem{larkin69}
A.~Larkin, Y.N. Ovchinnikov, Pis'ma Zh. Eksp. Teor. Fiz. \textbf{55}, 2262
  (1968)

\bibitem{Andree64}
A.~Andreev, Zh. Eksp. Teor. Fiz. \textbf{46}, 1823 (1964)

\bibitem{PhysRevB.25.4515}
G.E. Blonder, M.~Tinkham, T.M. Klapwijk, Phys. Rev. B \textbf{25}, 4515 (1982)

\bibitem{Pannetier2000}
B.~Pannetier, H.~Courtois, J. Low Temp. Phys. \textbf{118}, 599 (2000)

\bibitem{Klapwijk2004}
T.M. Klapwijk, Journal of Superconductivity \textbf{17}, 593 (2004)

\bibitem{PhysRevB.73.014503}
L.~Covaci, F.~Marsiglio, Phys. Rev. B \textbf{73}, 014503 (2006)

\bibitem{PhysRevB.81.014517}
A.M. Black-Schaffer, S.~Doniach, Phys. Rev. B \textbf{81}, 014517 (2010)

\bibitem{Triola19b}
C.~Triola, A.M. Black-Schaffer, Phys. Rev. B \textbf{100}, 024512 (2019)

\bibitem{Triola19}
C.~Triola, A.M. Black-Schaffer, arXiv:1905.00955  (2019)

\bibitem{PhysRevLett.16.453}
J.M. Rowell, W.L. McMillan, Phys. Rev. Lett. \textbf{16}, 453 (1966)

\bibitem{thanos2019}
A.~Tsintzis, A.M. Black-Schaffer, J.~Cayao, arXiv:1905.01171  (2019)

\bibitem{kuliksns}
I.O. Kulik, Sov. Phys. JETP \textbf{30}, 944 (1970)

\bibitem{Beenakker:92}
C.~Beenakker, \emph{Three "Universal" Mesoscopic Josephson Effects}, in
  \emph{Transport phenomena in mesoscopic systems: Proceedings of the 14th
  Taniguchi symposium, Shima, Japan, Nov. 10-14, 1991} (Springer-Verlag, 1992),
  p. 235

\bibitem{PhysRevLett.67.132}
A.~Furusaki, H.~Takayanagi, M.~Tsukada, Phys. Rev. Lett. \textbf{67}, 132
  (1991)

\bibitem{FURUSAKI1991299}
A.~Furusaki, M.~Tsukada, Solid State Commun. \textbf{78}, 299  (1991)

\bibitem{Triola18}
A.V. Balatsky, S.S. Pershoguba, C.~Triola, arXiv:1804.07244  (2018)

\bibitem{PhysRev.100.580}
G.~Dresselhaus, Phys. Rev. \textbf{100}, 580 (1955)

\bibitem{Rashba1960}
E.~Rashba, Sov. Phys. Solid. State \textbf{2}, 1109 (1960)

\bibitem{rashba84a}
Y.A. Bychkov, E.I. Rashba, Sov. Phys. JETP \textbf{39}, 78 (1984)

\bibitem{PhysRevLett.87.037004}
L.P. Gor'kov, E.I. Rashba, Phys. Rev. Lett. \textbf{87}, 037004 (2001)

\bibitem{PhysRevLett.92.027003}
E.~Bauer, G.~Hilscher, H.~Michor, C.~Paul, E.W. Scheidt, A.~Gribanov,
  Y.~Seropegin, H.~No\"el, M.~Sigrist, P.~Rogl, Phys. Rev. Lett. \textbf{92},
  027003 (2004)

\bibitem{PhysRevLett.92.097001}
P.A. Frigeri, D.F. Agterberg, A.~Koga, M.~Sigrist, Phys. Rev. Lett.
  \textbf{92}, 097001 (2004)

\bibitem{Reyren07}
N.~Reyren, S.~Thiel, A.D. Caviglia, L.F. Kourkoutis, G.~Hammerl, C.~Richter,
  C.W. Schneider, T.~Kopp, A.S. R\"{u}etschi, D.~Jaccard et~al., Science
  \textbf{317}, 1196 (2007)

\bibitem{doi:10.1143/JPSJ.76.051008}
S.~Fujimoto, J. Phys. Soc. Jpn. \textbf{76}, 051008 (2007)

\bibitem{PhysRevB.79.094504}
M.~Sato, S.~Fujimoto, Phys. Rev. B \textbf{79}, 094504 (2009)

\bibitem{PhysRevLett.113.227002}
X.~Liu, J.K. Jain, C.X. Liu, Phys. Rev. Lett. \textbf{113}, 227002 (2014)

\bibitem{0034-4885-80-3-036501}
M.~Smidman, M.B. Salamon, H.Q. Yuan, D.F. Agterberg, Rep. Prog. Phys.
  \textbf{80}, 036501 (2017)

\bibitem{0268-1242-11-8-009}
M.~Schultz, F.~Heinrichs, U.~Merkt, T.~Colin, T.~Skauli, S.~L{\o}vold,
  Semicond. Sci. Technol. \textbf{11}, 1168 (1996)

\bibitem{PhysRevLett.78.1335}
J.~Nitta, T.~Akazaki, H.~Takayanagi, T.~Enoki, Phys. Rev. Lett. \textbf{78},
  1335 (1997)

\bibitem{doi:10.1021/nl301325h}
D.~Liang, X.P. Gao, Nano Letters \textbf{12}, 3263 (2012)

\bibitem{Takase17}
K.~Takase, Y.~Ashikawa, G.~Zhang, K.~Tateno, S.~Sasaki, Sci. Rep. \textbf{7},
  930 (2017)

\bibitem{chang15}
W.~Chang, S.M. Albrecht, T.S. Jespersen, F.~Kuemmeth, P.~Krogstrup,
  J.~Nyg{\aa}rd, C.M. Marcus, Nat. Nanotech. \textbf{10}, 232 (2015)

\bibitem{Higginbotham}
A.P. Higginbotham, S.M. Albrecht, G.~Kirsanskas, W.~Chang, F.~Kuemmeth,
  P.~Krogstrup, T.S.J.J. Nyg{\aa}rd, K.~Flensberg, C.M. Marcus, Nat. Phys.
  \textbf{11}, 1017 (2015)

\bibitem{Krogstrup15}
P.~Krogstrup, N.L.B. Ziino, W.~Chang, S.M. Albrecht, M.H. Madsen, E.~Johnson,
  J.~Nyg{\aa}rd, C.M. Marcus, T.S. Jespersen, Nat. Mat. \textbf{14}, 400 (2015)

\bibitem{Deng16}
M.T. Deng, S.~Vaitiek\.{e}nas, E.B. Hansen, J.~Danon, M.~Leijnse, K.~Flensberg,
  J.~Nyg{\aa}rd, P.~Krogstrup, C.M. Marcus, Science \textbf{354}, 1557 (2016)

\bibitem{Albrecht16}
S.M. Albrecht, A.P. Higginbotham, M.~Madsen, F.~Kuemmeth, T.S. Jespersen,
  J.~Nyg{\aa}rd, , P.~Krogstrup, C.M. Marcus, Nature \textbf{531}, 206 (2016)

\bibitem{vaitienkenas17}
S.~Vaitiek\ifmmode~\dot{e}\else \.{e}\fi{}nas, M.T. Deng, J.~Nyg\aa{}rd,
  P.~Krogstrup, C.M. Marcus, Phys. Rev. Lett. \textbf{121}, 037703 (2018)

\bibitem{deng18}
M.T. Deng, S.~Vaitiek\ifmmode~\dot{e}\else \.{e}\fi{}nas, E.~Prada,
  P.~San-Jose, J.~Nyg\aa{}rd, P.~Krogstrup, R.~Aguado, C.M. Marcus, Phys. Rev.
  B \textbf{98}, 085125 (2018)

\bibitem{zhang16}
H.~Zhang, \"{O}nder G\"{u}l, S.~Conesa-Boj, K.~Zuo, V.~Mourik, F.K. de~Vries,
  J.~van Veen, D.J. van Woerkom, M.P. Nowak, M.~Wimmer et~al., Nat. Commun.
  \textbf{8}, 16025 (2017)

\bibitem{0957-4484-26-21-215202}
\"{O}nder G\"{u}l, D.J. van Woerkom, I.~van Weperen, D.~Car, S.R. Plissard,
  E.P.A.M. Bakkers, L.P. Kouwenhoven, Nanotechnology \textbf{26}, 215202 (2015)

\bibitem{Gazibegovic17}
S.~Gazibegovic, D.~Car, H.~Zhang, S.C. Balk, J.A. Logan, M.W.A. de~Moor, M.C.
  Cassidy, R.~Schmits, D.~Xu, G.~Wang et~al., Nature \textbf{548}, 434 (2017)

\bibitem{zhang18}
H.~Zhang, C.X. Liu, S.~Gazibegovic, D.~Xu, J.A. Logan, G.~Wang, N.~van Loo,
  J.D. Bommer, M.W. de~Moor, D.~Car et~al., Nature \textbf{556}, 74 (2018)

\bibitem{PhysRevMaterials.2.044202}
J.E. Sestoft, T.~Kanne, A.N. Gejl, M.~von Soosten, J.S. Yodh, D.~Sherman,
  B.~Tarasinski, M.~Wimmer, E.~Johnson, M.~Deng et~al., Phys. Rev. Materials
  \textbf{2}, 044202 (2018)

\bibitem{doi:10.1063/1.4971394}
S.T. Gill, J.~Damasco, D.~Car, E.P.A.M. Bakkers, N.~Mason, Appl. Phys. Lett.
  \textbf{109}, 233502 (2016)

\bibitem{gulonder}
O.~G\"{u}l, H.~Zhang, F.K. de~Vries, J.~van Veen, K.~Zuo, V.~Mourik,
  S.~Conesa-Boj, M.P. Nowak, D.J. van Woerkom, M.~Quintero-P\'{e}rez et~al.,
  Nano Letters \textbf{17}, 2690 (2017)

\bibitem{PhysRevB.81.184502}
L.~Santos, T.~Neupert, C.~Chamon, C.~Mudry, Phys. Rev. B \textbf{81}, 184502
  (2010)

\bibitem{PhysRevLett.95.146802}
C.L. Kane, E.J. Mele, Phys. Rev. Lett. \textbf{95}, 146802 (2005)

\bibitem{PhysRevLett.95.226801}
C.L. Kane, E.J. Mele, Phys. Rev. Lett. \textbf{95}, 226801 (2005)

\bibitem{Bernevig06}
B.A. Bernevig, T.L. Hughes, S.C. Zhang, Science \textbf{314}, 1757 (2006)

\bibitem{PhysRevLett.96.106802}
B.A. Bernevig, S.C. Zhang, Phys. Rev. Lett. \textbf{96}, 106802 (2006)

\bibitem{konig07}
M.~K\"{o}nig, S.~Wiedmann, C.~Br\"{u}ne, A.~Roth, H.~Buhmann, L.W. Molenkamp,
  Science \textbf{318}, 766 (2007)

\bibitem{Roth09}
A.~Roth, C.~Br\"{u}ne, H.~Buhmann, L.W. Molenkamp, J.~Maciejko, X.L. Qi, S.C.
  Zhang, Science \textbf{325}, 294 (2009)

\bibitem{brune2012}
C.~Br\"{u}ne, A.~Roth, H.~Buhmann, E.M. Hankiewicz, L.W. Molenkamp,
  J.~Maciejko, X.L. Qi, S.C. Zhang, Nat. Phys. \textbf{8}, 485 (2012)

\bibitem{Nowack2013}
K.C. Nowack, E.M. Spanton, M.~Baenninger, M.~K\"{o}nig, J.R. Kirtley,
  B.~Kalisky, C.~Ames, P.~Leubner, C.~Br\"{u}ne, H.~Buhmann et~al., Nat. Mater.
  \textbf{12}, 787 (2013)

\bibitem{PhysRevLett.100.236601}
C.~Liu, T.L. Hughes, X.L. Qi, K.~Wang, S.C. Zhang, Phys. Rev. Lett.
  \textbf{100}, 236601 (2008)

\bibitem{PhysRevLett.107.136603}
I.~Knez, R.R. Du, G.~Sullivan, Phys. Rev. Lett. \textbf{107}, 136603 (2011)

\bibitem{RevModPhys.82.3045}
M.Z. Hasan, C.L. Kane, Rev. Mod. Phys. \textbf{82}, 3045 (2010)

\bibitem{RevModPhys.83.1057}
X.L. Qi, S.C. Zhang, Rev. Mod. Phys. \textbf{83}, 1057 (2011)

\bibitem{doi:10.1002/pssb.201248385}
G.~Tkachov, E.M. Hankiewicz, Phys. Status Solidi (b) \textbf{250}, 215 (2013)

\bibitem{Ando13}
Y.~Ando, J. Phys. Soc. Jpn. \textbf{82}, 102001 (2013)

\bibitem{Sato_2017}
M.~Sato, Y.~Ando, Reports on Progress in Physics \textbf{80}, 076501 (2017)

\bibitem{Yacoby14}
S.~Hart, H.~Ren, T.~Wagner, P.~Leubner, M.~M\"{u}hlbauer, C.~Br\"{u}ne,
  H.~Buhmann, L.W. Molenkamp, A.~Yacoby, Nat. Phys. \textbf{10}, 638 (2014)

\bibitem{vlad15}
V.S. Pribiag, A.J.A. Beukman, F.~Qu, M.C. Cassidy, C.~Charpentier,
  W.~Wegscheider, L.P. Kouwenhoven, Nat. Nanotech. \textbf{10}, 593 (2015)

\bibitem{PhysRevB.82.081303}
P.~Adroguer, C.~Grenier, D.~Carpentier, J.~Cayssol, P.~Degiovanni, E.~Orignac,
  Phys. Rev. B \textbf{82}, 081303 (2010)

\bibitem{PhysRevLett.109.186603}
I.~Knez, R.R. Du, G.~Sullivan, Phys. Rev. Lett. \textbf{109}, 186603 (2012)

\bibitem{PhysRevB.81.241310}
T.D. Stanescu, J.D. Sau, R.M. Lutchyn, S.~Das~Sarma, Phys. Rev. B \textbf{81},
  241310 (2010)

\bibitem{PhysRevB.88.075401}
G.~Tkachov, E.M. Hankiewicz, Phys. Rev. B \textbf{88}, 075401 (2013)

\bibitem{Bocquillon17}
E.~Bocquillon, R.S. Deacon, J.~Wiedenmann, P.~Leubner, T.M. Klapwijk,
  C.~Br\"{u}ne, K.~Ishibashi, , H.~Buhmann, L.W. Molenkamp, Nat. Nanotech.
  \textbf{12}, 137 (2017)

\bibitem{PhysRevB.86.214515}
C.T. Olund, E.~Zhao, Phys. Rev. B \textbf{86}, 214515 (2012)

\bibitem{PSSB:PSSB201248385}
G.~Tkachov, E.M. Hankiewicz, Phys. Status Solidi B \textbf{250}, 215 (2013)

\bibitem{PhysRevB.74.180503}
C.~Benjamin, Phys. Rev. B \textbf{74}, 180503 (2006)

\bibitem{PhysRevB.90.205435}
B.H. Wu, W.~Yi, J.C. Cao, G.C. Guo, Phys. Rev. B \textbf{90}, 205435 (2014)

\bibitem{PhysRevLett.93.197003}
D.~Beckmann, H.B. Weber, H.~v.~L\"ohneysen, Phys. Rev. Lett. \textbf{93},
  197003 (2004)

\bibitem{PhysRevLett.95.027002}
S.~Russo, M.~Kroug, T.M. Klapwijk, A.F. Morpurgo, Phys. Rev. Lett. \textbf{95},
  027002 (2005)

\bibitem{strunz19}
J.~Strunz, J.~Wiedenmann, C.~Fleckenstein, L.~Lunczer, W.~Beugeling, V.L.
  M\"{u}ller, P.~Shekhar, N.T. Ziani, S.~Shamim, J.~Kleinlein et~al.,
  arXiv:1905.08175  (2019)

\bibitem{majorana}
E.~Majorana, Nuovo Cimento pp. 171--184 (1937)

\bibitem{PhysRevB.86.085408}
J.~Klinovaja, D.~Loss, Phys. Rev. B \textbf{86}, 085408 (2012)

\bibitem{PhysRevB.86.180503}
E.~Prada, P.~San-Jose, R.~Aguado, Phys. Rev. B \textbf{86}, 180503 (2012)

\bibitem{PhysRevB.86.220506}
S.~Das~Sarma, J.D. Sau, T.D. Stanescu, Phys. Rev. B \textbf{86}, 220506 (2012)

\bibitem{PhysRevB.87.024515}
D.~Rainis, L.~Trifunovic, J.~Klinovaja, D.~Loss, Phys. Rev. B \textbf{87},
  024515 (2013)

\bibitem{PhysRevB.84.195442}
T.P. Choy, J.M. Edge, A.R. Akhmerov, C.W.J. Beenakker, Phys. Rev. B
  \textbf{84}, 195442 (2011)

\bibitem{magnatoms}
D.J. Choi, N.~Lorente, J.~Wiebe, K.~von Bergmann, A.F. Otte, A.J. Heinrich,
  arXiv:1904.09941  (2019)

\bibitem{PhysRevB.92.121404}
Z.~Huang, P.~W\"olfle, A.V. Balatsky, Phys. Rev. B \textbf{92}, 121404 (2015)

\bibitem{lutchyn16}
S.P. Lee, R.M. Lutchyn, J.~Maciejko, Phys. Rev. B \textbf{95}, 184506 (2017)

\bibitem{Takagi18}
D.~Takagi, S.~Tamura, Y.~Tanaka, arXiv:1809.09324  (2018)

\bibitem{tamura18}
S.~Tamura, S.~Hoshino, Y.~Tanaka, Phys. Rev. B \textbf{99}, 184512 (2019)

\bibitem{PhysRevB.87.104513}
Y.~Asano, Y.~Tanaka, Phys. Rev. B \textbf{87}, 104513 (2013)

\bibitem{PhysRevB.70.012507}
Y.~Tanaka, S.~Kashiwaya, Phys. Rev. B \textbf{70}, 012507 (2004)

\bibitem{PhysRevB.92.014513}
X.~Liu, J.D. Sau, S.~Das~Sarma, Phys. Rev. B \textbf{92}, 014513 (2015)

\bibitem{PhysRevB.53.9371}
Y.~Tanaka, S.~Kashiwaya, Phys. Rev. B \textbf{53}, 9371 (1996)

\bibitem{Kwon2004}
H.J. Kwon, K.~Sengupta, V.M. Yakovenko, Eur. Phys. J. B \textbf{37}, 349 (2004)

\bibitem{PhysRevB.90.045118}
J.~Klinovaja, D.~Loss, Phys. Rev. B \textbf{90}, 045118 (2014)

\bibitem{Baba_2018}
S.~Baba, C.~J\"{u}nger, S.~Matsuo, A.~Baumgartner, Y.~Sato, H.~Kamata, K.~Li,
  S.~Jeppesen, L.~Samuelson, H.Q. Xu et~al., New J. Phys. \textbf{20}, 063021
  (2018)

\bibitem{PhysRevB.95.174516}
O.~Kashuba, B.~Sothmann, P.~Burset, B.~Trauzettel, Phys. Rev. B \textbf{95},
  174516 (2017)

\bibitem{DushkoMZM}
D.~Kuzmanovski, A.M. Black-Schaffer, J.~Cayao, in preparation.

\bibitem{PhysRevB.99.024506}
N.V. Gnezdilov, Phys. Rev. B \textbf{99}, 024506 (2019)

\bibitem{PhysRevLett.70.3339}
S.~Sachdev, J.~Ye, Phys. Rev. Lett. \textbf{70}, 3339 (1993)

\bibitem{kitaevsyk}
A.~Kitaev, a simple model of quantum holography,? talks at KITP, April 7, 2015
  and May 27, 2015, \url{http://online.kitp.ucsb.edu/online/entangled15/}

\bibitem{PhysRevB.96.121119}
A.~Chew, A.~Essin, J.~Alicea, Phys. Rev. B \textbf{96}, 121119 (2017)

\bibitem{gaplesssc}
A.~Abrikosov, L.~Gor'kov, Sov. Phys. JETP \textbf{12}, 1243 (1961)

\bibitem{PhysRevLett.9.315}
F.~Reif, M.A. Woolf, Phys. Rev. Lett. \textbf{9}, 315 (1962)

\bibitem{PhysRevLett.10.96}
J.C. Phillips, Phys. Rev. Lett. \textbf{10}, 96 (1963)

\end{thebibliography}

\end{document}